\documentclass[journal]{IEEEtran}
\usepackage{xcolor,soul,framed} 
\usepackage{cite}
\usepackage{amsmath,amssymb,amsfonts}
\usepackage{graphicx}
\usepackage{textcomp}
\usepackage{dcolumn}
\usepackage{bm}
\usepackage{subfigure}
\usepackage{pict2e}
\usepackage{epstopdf}
\usepackage{arydshln}
\usepackage{gensymb}
\usepackage{enumitem}
\usepackage[mathscr]{euscript}
\usepackage{dblfloatfix} 

\definecolor{bluerevision}{RGB}{0,112,192}

\newcommand{\E}{\mathrm{e}}
\newcommand{\jj}{\mathrm{j}}

\makeatletter
\def\hlinewd#1{%
\noalign{\ifnum0=`}\fi\hrule \@height #1 \futurelet
\reserved@a\@xhline}
\makeatother

\makeatletter
 \let\old@ps@headings\ps@headings
 \let\old@ps@IEEEtitlepagestyle\ps@IEEEtitlepagestyle
 \def\confheader#1{%
 \def\ps@headings{%
 \old@ps@headings%
 \def\@oddhead{\strut\hfill#1\hfill\strut}%
 \def\@evenhead{\strut\hfill#1\hfill\strut}%
 }%
 \def\ps@IEEEtitlepagestyle{%
 \old@ps@IEEEtitlepagestyle%
 \def\@oddhead{\strut\hfill#1\hfill\strut}%
 \def\@evenhead{\strut\hfill#1\hfill\strut}%
 }%
 \ps@headings%
 }
 \makeatother
\confheader{%
This work has been accepted for publication in \textit{IEEE Transactions on Wireless Communications}. DOI: 10.1109/TWC.2024.3377539
}

\begin{document}

\title{Joint Ultra-wideband Characterization of Azimuth, Elevation and Time of Arrival with Toric Arrays}

\author{Alejandro Ramírez-Arroyo, Antonio Alex-Amor, Rubén Medina, Pablo Padilla, and Juan F. Valenzuela-Valdés 
\thanks{This work has been supported by grant PID2020-112545RB-C54 funded by MCIN/AEI/10.13039/501100011033. It has also been supported by grants PDC2022-133900-I00, PDC2023-145862-I00, TED2021-129938B-I00 and TED2021-131699B-I00 funded by MCIN/AEI/10.13039/501100011033 and by the European Union NextGenerationEU/PRTR; and by Universidad de Granada through grant PPJIB2022-05; and in part by the Predoctoral Grant FPU19/01251 and FPU19/04085. (\textit{Corresponding author: Alejandro Ramírez-Arroyo}.)}
\thanks{Alejandro Ramírez-Arroyo, Pablo Padilla and Juan F. Valenzuela-Valdés are with the Department of Signal Theory, Telematics and Communications, Research Centre for Information and Communication Technologies (CITIC-UGR), Universidad de Granada, 18071 Granada, Spain (e-mail: alera@ugr.es; pablopadilla@ugr.es;
juanvalenzuela@ugr.es).}
\thanks{Antonio Alex-Amor is with the Department of Information Technologies, Universidad San Pablo-CEU, CEU Universities,  Campus Montepríncipe, 28668 Boadilla del Monte (Madrid), Spain  (e-mail: antonio.alexamor@ceu.es).}
\thanks{Rubén Medina is with the Department of Mathematical Analysis, Universidad de Granada (UGR), 18071 Granada, Spain (e-mail: rubenmedina@ugr.es).}
}

\markboth{Ramírez-Arroyo \MakeLowercase{\textit{et al.}}: Joint Ultra-wideband Characterization of Azimuth, Elevation and Time of Arrival with Toric Arrays}%
{Ramírez-Arroyo \MakeLowercase{\textit{et al.}}: Joint Characterization of Azimuth, Elevation and Time of Arrival with Toric Arrays}

\IEEEpeerreviewmaketitle

\maketitle


\begin{abstract}
In this paper, we present an analytical framework for the joint characterization of the 3D direction of arrival (DoA), i.e., azimuth and elevation components, and time of arrival (ToA) in multipath environments. The analytical framework is based on the use of nearly frequency-invariant beamformers (FIB) formed by toric arrays. The frequency response of the toric array is expanded as a series of phase modes, which leads to azimuth--time and elevation--time diagrams from which the 3D DoA and the ToA of the incoming waves can be extracted over a wide bandwidth. Firstly, we discuss some practical considerations, advantages and limitations of using the analytical method. Subsequently, we perform a parametric study to analyze the influence of the method parameters on the quality of the estimation. The method is tested in single-path and multipath mm-wave environments over a large bandwidth. The results show that the proposed method improves the quality of the estimation, i.e., decreases the level of the artifacts, compared to other state-of-art FIB approaches based on the use of single/concentric circular and elliptical arrays.
\end{abstract}


\newcommand*{\bigs}[1]{\vcenter{\hbox{\scalebox{2}[8.2]{\ensuremath#1}}}}

\newcommand*{\bigstwo}[1]{\vcenter{\hbox{ \scalebox{1}[4.4]{\ensuremath#1}}}}

\section{Introduction}

\IEEEPARstart{I}{mproving} the performance of wireless links requires a proper characterization and knowledge of multiple channel parameters: direction of arrival (DoA), time of arrival (ToA), delay spread, path loss, and $K$ factor, among others \cite{KPI_2, parameters_1}. With knowledge of the channel parameters, different scenarios can be effectively distinguished \cite{Alex_IA}, even recreated and emulated through the use of post-processing techniques based on the creation/removal of reflections in the communication channel \cite{TimeGating, TimeGating2}. This is fundamental in today's interconnected society, considering the huge variety of propagation environments associated with different communication scenarios: 5G/6G mobile \cite{mobile1, mobile3}, RIS-aided \cite{RIS0, RIS2}, industrial \cite{factory1, factory3}, and vehicle-to-everything \cite{v2x1, t2t, uav, s2s, uav_model} networks. 

Among all the channel parameters, DoA and ToA are among the most noteworthy. A joint estimation of DoA and ToA is essential because current communication scenarios suffer from temporal variations of the physical properties of the channel. Accurate DoA and ToA information can account for the changes occurring in the channel.  In the sub-6 GHz regime, DoA and ToA estimation techniques have been widely explored in the literature \cite{twodecades}. Most of the applied techniques are of narrowband nature \cite{narrowbandDoA0, DoA1, esprit,  maximum_likelihood, narrowbandDoA1, narrowbandDoA2, mimo_model}, which generally limits their use for modern broadband applications in the mm-wave frequency range. As a consequence, new efficient wideband alternatives are being sought that feature DoA and ToA simultaneously. For instance, narrowband approximations can be extended towards wideband applications through the use of time-delay beamformers \cite{time_delay_beamformer}. These approaches propose a frequency-dependent phase shift, which is applied on a narrowband decomposition of the wideband communication channel. Thus, the combination of narrowband nature approximations and frequency-dependent phase shifters leads to wideband approaches. One step beyond, the whole wideband channel can be considered through the design and implementation of \emph{nearly frequency-invariant} beamformers (FIB), which has proven to be a suitable alternative for (ultra)-wideband DoA estimation \cite{fib1, fib2, fib3}. The objective of FIB is to parameterize the array coefficients so that the spectral and spatial dependences can be treated independently \cite{fib4}. Previous implementations of FIB have ranged from the use of one-dimensional (1D) arrays \cite{fib5} to two-dimensional (2D) configurations based on the use of circular arrays \cite{phase_mode2002, phase_mode2007, phase_mode2008, Fan_2019}.  Recently, the use of FIB was extended by the authors to include elliptical geometries \cite{AlexElipses2023, AlexEuCAP2023}. This is a generalization of previous approaches based on circular geometries, as circular and linear arrays are subcases included in the more general elliptical arrays. However, many of the DoA and ToA estimation methods are only capable to accurately estimate the time of arrival and one of the two spatial DoA components, i.e., either azimuth or elevation, but not both at the same time \cite{phase_mode2002, phase_mode2007, phase_mode2008, AlexElipses2023, AlexEuCAP2023}. In this regard, it is not so common to find analytical approaches that accurately estimate azimuth, elevation and time of arrival in one go.

\begin{figure*}[t]
	\centering
	\subfigure[]{\includegraphics[width=0.95\columnwidth]{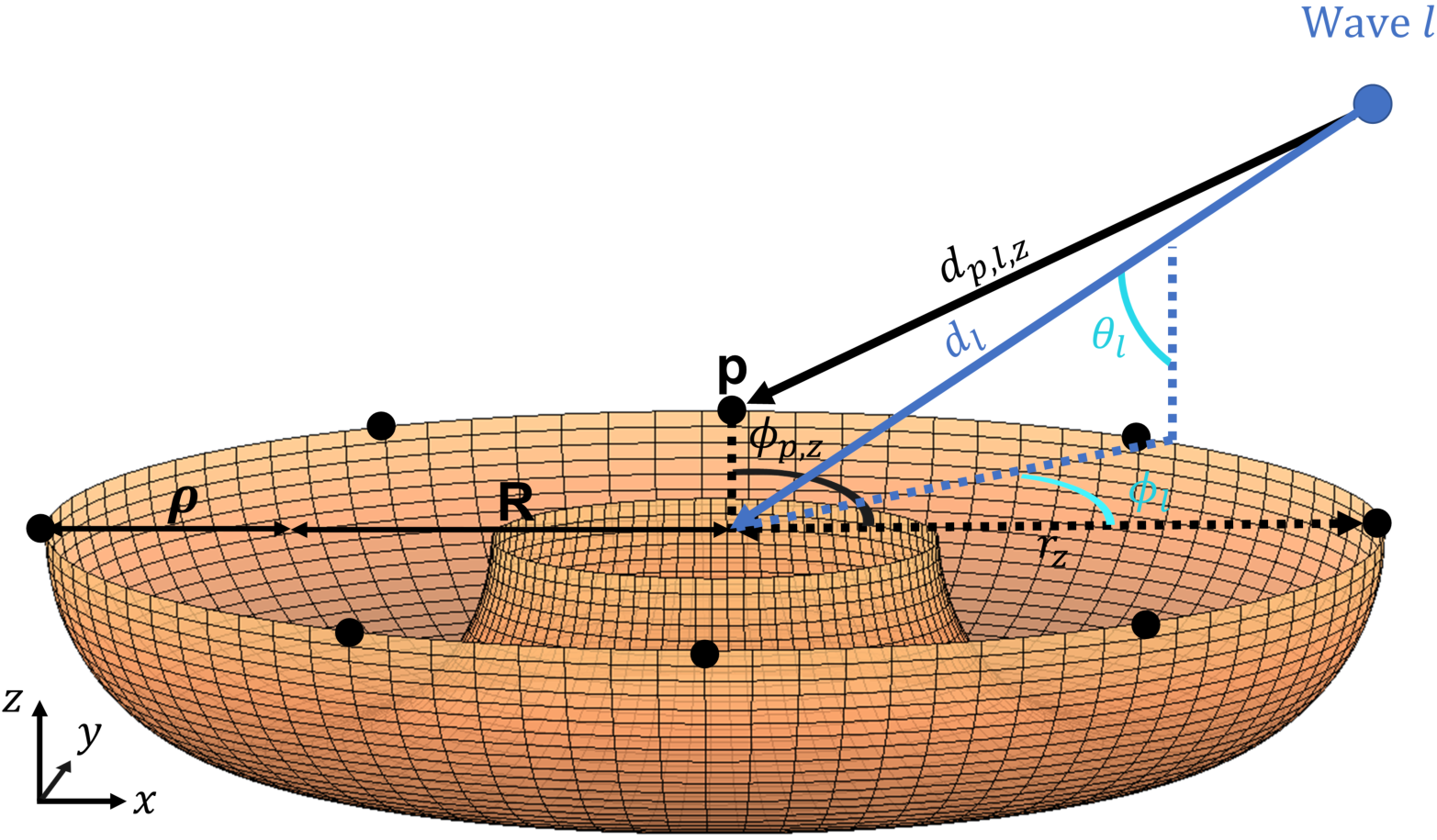}
	} 
\subfigure[]{\includegraphics[width= 0.95\columnwidth]{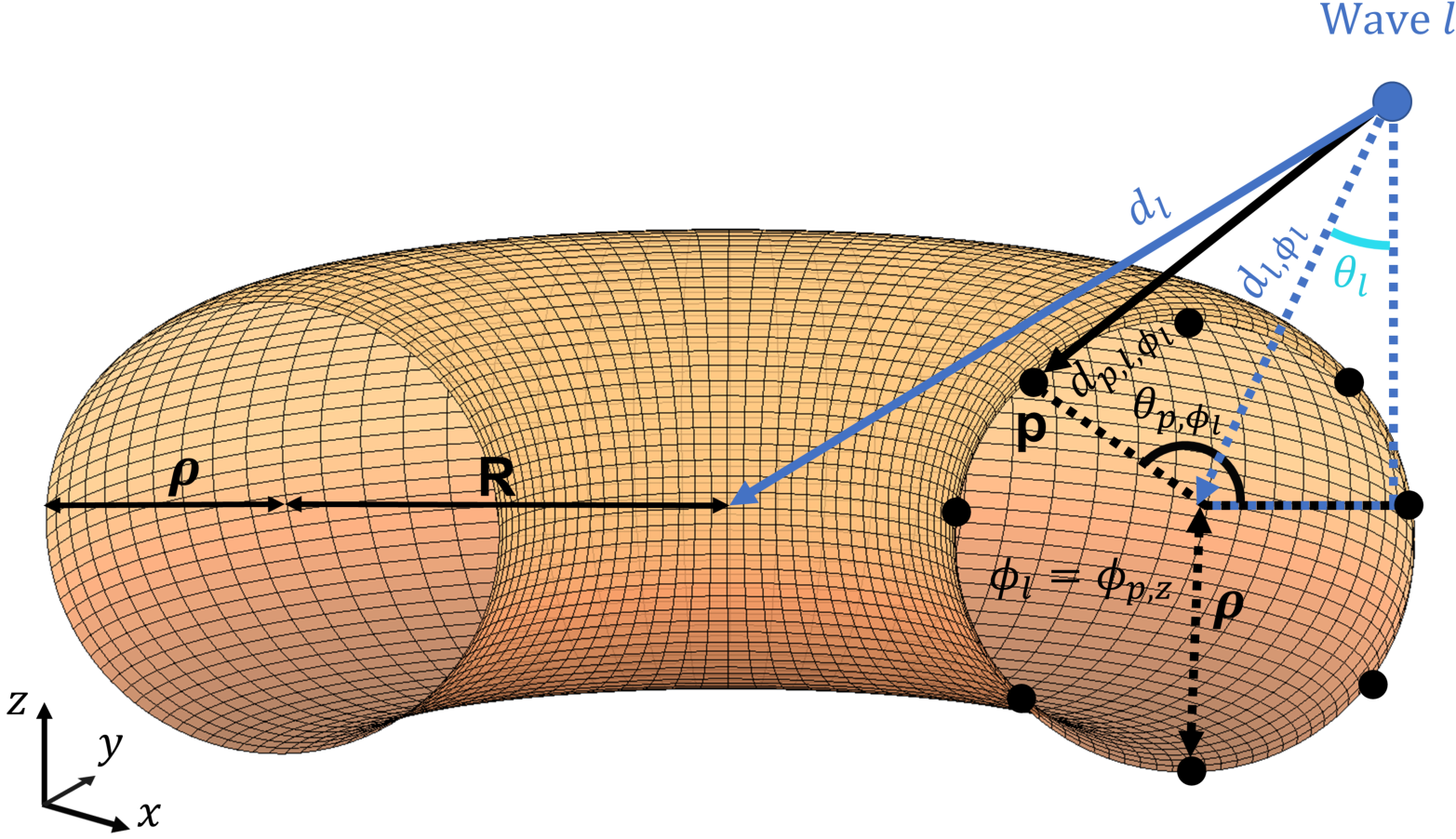}
	}
	\caption{Torus geometry:  Representation of the circles arranged in (a) the XY plane, and (b) $\Phi_l$Z plane.} 
	\label{esquema_toros}
\end{figure*}

In this paper, we present a novel estimation method for joint 3D DoA (azimuth, elevation) and ToA characterization. The technique is based on the use of nearly frequency-invariant toric arrays. Following a similar rationale than in \cite{phasemode_original, phase_mode2002, phase_mode2008, phase_mode2007, Zhang_2017, Fan_2019, AlexElipses2023, AlexEuCAP2023}, the multipath frequency response, acquired in the different spatial points that are part of the toric array, can be expanded as a series of phase modes and a preselected frequency-dependent filter. This leads to diagrams in the azimuth--time (AoA--ToA) and elevation--time (EoA--ToA) domains, from which the direction and time of arrival of the incoming waves can be accurately estimated in a wideband range of frequencies in either single-path or multipath environments. This is a remarkable feature of the proposed method, as a joint 3D-DoA and ToA characterization is rarely found in the literature. Moreover, the accuracy in the joint DoA and ToA estimation is improved with respect to previous state-of-art frequency-invariant beamformers. This is essentially due to the geometry of the torus. The geometrical disposal of spatial samples in a toric array efficiently exploits the arrangement for an optimal estimation. Within a torus, multiple concentric circular arrays can be defined in horizontal planes, thus improving the estimation of the azimuth angle. Then, the optimal vertical plane, which includes a single circular array, is selected for a precise estimation of the elevation angle and time of arrival. Finally, it is worth mentioning that the proposed technique works efficiently in under-sampling conditions, which can be used to reduce processing time and the number of samples employed. Thus, the main contributions of this work are summarized as:

\begin{enumerate}[label=(\roman*)]

    \item Development of the theoretical framework for the use of frequency-invariant beamformers in three-dimensional geometries. The distributions of samples in three-dimensional spaces provided by toric arrays allow the development of expressions that lead to a three-dimensional characterization of the direction of arrival (azimuth and elevation) and time of arrival.
    
    \item Method validation through simulation in the mm-wave frequency range. This approach can be employed in wideband single-path and multi-path channels by decoupling the spatial and temporal domains given the channel frequency response of each spatial sample.

    \item Parametric analysis of the toric geometry. A study of the geometrical parameters of the torus given the above framework is performed to define the optimal region of operation of the method.
    
\end{enumerate}   

\noindent These contributions open up the possibility of angular and temporal characterization of a 3D multipath scenario. This is achieved by means of the frequency-invariant beamforming framework proposed for toric arrangements throughout the paper. Moreover, as it will be detailed later, the proposed framework improves the quality of the DoA and ToA estimation compared to other state-of-the-art FIB approaches based on the use of circular and elliptical arrays.

The document is organized as follows. Section II presents the mathematical framework for the joint characterization of the azimuth, elevation and time of arrival. It also discusses some practical considerations, advantages and limitations of the method. Section III illustrates some numerical examples, including single-path and multipath scenarios, in order to validate the proposed theoretical framework. A parametric study on how the main involved parameters affect the performance of the method is also carried out. Finally, general conclusions are drawn in Section IV.


\section{\label{sec:Theory} Theoretical Framework}

A torus is defined as a closed surface formed by the Cartesian product of two circles. Parametrically, it can be defined as:

\begin{equation} \label{ecuacion_toro_xyz}
\begin{aligned}
& x (\phi, \theta) = (R+\rho \sin \theta) \cos \phi \\
& y (\phi, \theta) = (R+\rho \sin \theta) \sin \phi, \\
& z (\phi, \theta) = -\rho \cos \theta
\end{aligned}
\end{equation}

\noindent where $R$ is the distance from the torus center to the tube center, and $\rho$ is the tube radius. $\phi$ is the azimuth angle, which represents the rotation around the axis of revolution. $\theta$ is the elevation angle, i.e., the rotation angle around the tube. $R > \rho$ is considered in order to ensure a ring torus shape. 

Due to the nature of this geometry, circles can be generated in the XY plane given a specific $z$ height. Since $z (\phi, \theta)$ only depends on $\theta$ for any value of $\phi$, $\theta$ can be directly substituted in $x (\phi, \theta)$  and $y (\phi, \theta)$ as $\sin(\theta) = \sin(\arccos(-z/\rho)) = \sqrt{1-(-z/\rho)^2)}$. This expression can be further simplified, leading (\ref{ecuacion_toro_xyz}) to:

\begin{equation} \label{ecuacion_toro_xyz_simp}
\begin{aligned}
& x (z, \phi) = (R \pm \sqrt{\rho^2 - z^2}) \cos \phi \\
& y (z, \phi) = (R \pm \sqrt{\rho^2 - z^2}) \sin \phi. \\
\end{aligned}
\end{equation}

\begin{figure}[t]
	\centering
	\includegraphics[width= 1\columnwidth]{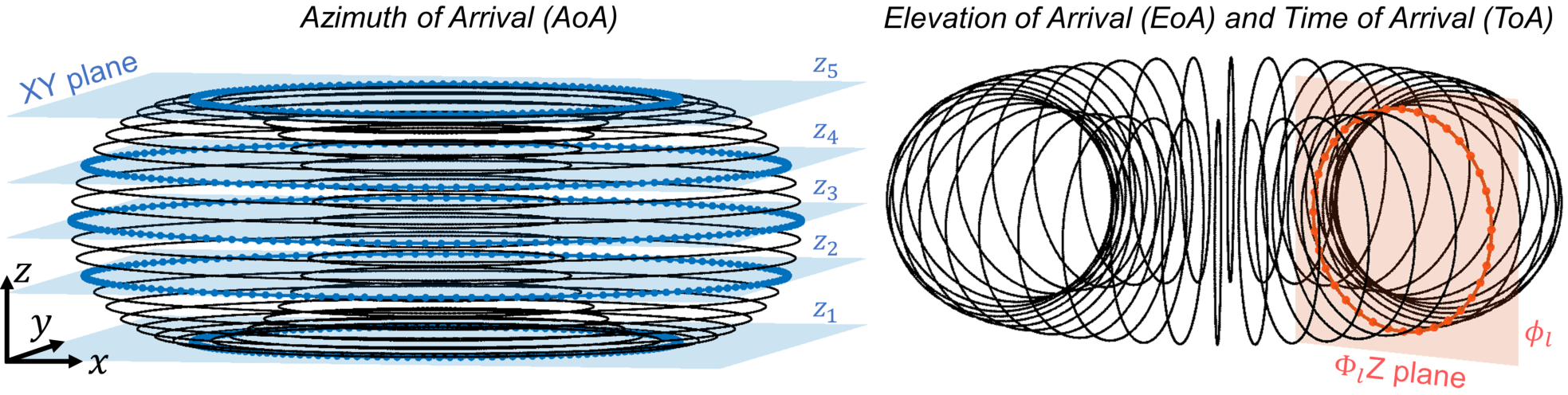}
	\caption{Planes where the circular arrays are placed on the torus. $P$ circular arrays can be found in the XY plane given several heights $z$ (left panel). For visualization purposes, only five circles are marked in blue in the XY plane.  A single circular array is located in the $\Phi_l$Z plane (right panel). The $P$ circular arrays in the XY plane and the single array in the $\Phi_l$Z plane allow us to estimate the AoA and the pair \{EoA, ToA\}, respectively.} 
	\label{esquema_cortes}
\end{figure}

\noindent An example of circumference in this XY plane is illustrated in Fig. \ref{esquema_toros}(a) when $z = 0$. Therefore, several circumferences can be arranged given the horizontal XY plane for several $z$ values. Equivalently, two circumferences are found in the vertical plane formed by the Z-axis and a given angle $\phi$. Fig.~\ref{esquema_toros}(b) shows this plane when $\phi~=~\phi_l$, which represents a smart selection of the plane given the $\phi\textrm{-axis}$. Mathematically, the $\Phi_l$Z~plane is defined by the line that lies in the Cartesian \mbox{$z$-axis} with unit direction vector $\mathbf{u} = \hat{\mathbf{z}}$, and the line that crosses the origin \{0,0,0\} and the coordinate \{$\cos (\phi_l), \sin (\phi_l), 0$\} with unit direction vector $\mathbf{v} = \cos(\phi_l) \hat{\mathbf{x}} + \sin(\phi_l) \hat{\mathbf{y}} $. Thus, the $\Phi_l$Z~plane is defined by the plane equation \mbox{$\sin(\phi_l) x -  \cos(\phi_l) y = 0$.} Some examples of these arrangements are illustrated in Fig.~\ref{esquema_cortes}. These distributions will be analyzed in detail in later sections. These properties of the torus to define circumferences in several planes will be used to accurately estimate the 3D DoA (azimuth $\phi$ and elevation $\theta$), as well as the ToA $\tau$ in multipath environments.

Now, let us assume an incident spherical wave $l$ characterized by the azimuth angle $\phi_l$, elevation $\theta_l$ and time of arrival $\tau_l$. This signal impinges on a torus formed by $P$ samples on each circumference arranged in the XY plane, and $P$ samples on each circumference of the $\Phi_l$Z plane, for a total of $P^{2}$ spatial samples. The estimation of the three previous parameters is performed in two steps: (i)~The first one takes advantage of the circumferences located in the XY plane to make an accurate estimation of $\phi_l$. (ii)~The second step makes use of the previous value of $\phi_l$ to accurately estimate $\theta_l$ and $\tau_l$ given the $\Phi_l$Z plane.

\subsection{Azimuth of Arrival (AoA)}

In the first step, the frequency response at the center of the circumferences lying in the XY plane is expressed as

\begin{equation} \label{H_lz}
H_{l,z}(f)=\kappa_l \mathrm{e}^{\mathrm{j} 2 \pi f \tau_{l,z}} =\kappa_l \mathrm{e}^{\mathrm{j} 2 \pi f (\tau_{l} + \tau_{z})},
\end{equation}

\noindent where $\kappa_l$ stands for the complex amplitude for the $l$-th wave and $\tau_{l,z}$ for the delay of the $l$-th wave for a given height $z$. The term $\tau_{l,z}$ can be expressed as $\tau_{l,z} = \tau_{l} + \tau_{z}$, where $\tau_{l}$ is the delay characterized at the center of the torus ($z = 0$) and $\tau_z$ is an additional delay introduced by the height $z$. According to the torus geometry, the largest difference between $\tau_z$ values is given by $2\rho/c$, which is the diameter of the torus tube divided by the wave propagation speed, i.e., the speed of light $c$. This is the case for $\theta_l = 0\degree$ and $\theta_l = 180\degree$. For $\theta_l = 90\degree$, the incident wave impinges on the different circles at the same time, so the difference between values of $\tau_z$ becomes zero. Generally, it is satisfied that $\tau_l \gg 2\rho/c$, so we can assume that $\tau_{l,z} \approx \tau_l$. This effect generates a slight temporal spread in the time domain estimation, which will be analyzed in the following sections. Regardless of this fact, note that there is no negative effect on the estimation since the ToA is not yet estimated in this step.

For each circle at a height $z$, $H_{p,l,z}(f)$ reads as the channel frequency response acquired at the $p$-th sample. Since the spatial samples are uniformly distributed around the position of $H_{l,z}(f)$, $H_{p,l,z}(f)$ is given by

\begin{equation} \label{Hplz}
H_{p, l, z}(f)=\left(\frac{d_{l,z}}{d_{p, l, z}}\right)^{\gamma/2} H_{l,z}(f)\, \E^{\jj 2 \pi f \Delta d_{p, l, z} / c}\, ,
\end{equation}

\noindent where $d_{l,z} = c\,\tau_{l,z}$, and $d_{p,l,z}$ is the distance traveled by the wave to reach the $p$-th spatial sample. Therefore, the term $\left(d_{l,z}/d_{p, l, z}\right)^{\,\,\gamma/2}$ is the attenuation factor given the distance between the torus center at a height $z$, and the $p$-th sample for a path loss exponent~$\gamma$. The complex exponential term stands for the phase shift introduced by the additional distance to the center of the torus, where

\begin{equation} \label{Delta_d_plz}
\Delta d_{p,l,z} = d_{l,z} - d_{p,l,z}.
\end{equation}

\noindent According to the torus geometry, the previous term can be expanded as

\begin{equation} \label{delta_dplz_v2}
\Delta d_{p,l,z} = d_{l,z} - \sqrt{d_{l,z}^2+r{_z}^2-2 d_{l,z} r_z \sin (\theta_{l}) \cos \left(\phi_{l}-\phi_{p,z}\right)}
\end{equation}
where $\phi_{p,z}$ is the azimuth angle for the $p$-th sample located at a height $z$, which is evenly spaced in the angular domain $\phi \in [0,2\pi)$. The term $r_z$ stands for the radius of the circles deployed at different heights. From (\ref{ecuacion_toro_xyz_simp}), it is defined as

\begin{equation}
r_{z} = R \pm \sqrt{\rho^2 - z^2}.
\end{equation}

\noindent Through the expansion in Taylor series, (\ref{delta_dplz_v2}) can be approximated as

\begin{equation} \label{taylor_series}
\Delta d_{p,l,z} \approx r_z \sin (\theta_{l}) \cos (\phi_{l}-\phi_{p,z}).
\end{equation}

\noindent which allows us to simplify (\ref{Hplz}) to

\begin{equation} \label{Hplz_v2}
H_{p, l, z}(f)=\left(\frac{d_{l,z}}{d_{p, l, z}}\right)^{\gamma/2} H_{l,z}(f)\, \E^{\jj 2 \pi f r_z \sin (\theta_{l}) \cos (\phi_{l}-\phi_{p,z}) / c}\, ,
\end{equation}

As a starting point, a first approach considers $\sin(\theta_l) = 1$ by fixing elevation incidence angle $\theta_l =\nolinebreak90\degree$. This approximation is fundamental since it will later allow the decoupling of the azimuth angular term $\phi_l$ from the frequency term $f$ through the Jacobi-Anger identity. Note that this approach is feasible because FIBs that have been developed to be robust over multiple elevation angles will be applied in later steps \cite{Zhang_2017}, while the accurate estimation of the elevation angle $\theta_l$ is performed in Section II.B. Additionally, $d_{l,z}/d_{p,l,z} \approx 1$ for $d_{l,z} \gg \Delta d_{p,l,z}$, i.e. the wave source is far away from the toric array (plane wave approximation). Note that this approach is necessary for the development of the theoretical framework throughout this section. However, numerical simulations and experiments performed with circular and elliptical arrays have shown that the characterization of the propagation channel is feasible, even in the near-field region, i.e. considering spherical wave propagation \cite{AlexElipses2023,AlexEuCAP2023}. After these two considerations, (\ref{Hplz_v2}) can be simplified to
\begin{equation} \label{Hplz_simplificada}
H_{p, l, z}(f)= H_{l,z}(f)\, \E^{\jj 2 \pi f r_z \cos (\phi_{l}-\phi_{p,z}) / c}\, .
\end{equation}

\noindent As observed in \eqref{Hplz_simplificada}, phase and frequency components are linked in the complex exponential term. The Jacobi-Anger identity, given by \cite{Jacobi-Anger}

\begin{equation} \label{JacobiAnger_ecuacion}
\E^{\jj \upsilon \cos \alpha} = \sum_{n=-\infty}^{\infty} \jj^n J_n(\upsilon) \E^{\jj n \alpha} \, ,
\end{equation}
\noindent allows us to decouple phase ($\phi_l$) and frequency ($f$) terms, extending (\ref{Hplz_simplificada}) to
\begin{equation}
H_{p, l, z}(f)=H_{l,z}(f) \sum_{n=-\infty}^{\infty} \mathrm{j}^n J_{n, z}\left(\frac{2 \pi f r_z}{c}\right) \mathrm{e}^{\mathrm{j} n\left(\phi_l-\phi_{p,z}\right)}.
\end{equation}

\noindent The former step opens up the possibility of a solution based on nearly frequency-independent beamformers (FIBs). Note that $J_{n, z}\left(\cdot\right)$ is the Bessel function of the first kind of order $n$ for the circular array with radius $r_{z}$.

The former expression can be projected onto basis functions of the form $\E^{\jj m\phi_{p,z}}$ that excite the frequency response $H_{p,l,z}(f)$ when $\phi_{p,z} = \phi_l$. This method is known as phase-mode expansion \cite{phasemode_original, phase_mode2002}, which provides the phase-mode domain $\widehat{H}_{m,l,z}(f)$:
\begin{equation}
\label{eq13}
\begin{aligned}
&\widehat{H}_{m, l, z}(f) \\
&=\frac{1}{P} \sum_{p=0}^{P-1} H_{p, l, z}(f)\, \E^{\jj m \phi_{p,z}}  \\
&=H_{l,z}(f) \sum_{n=-\infty}^{+\infty} \mathrm{j}^n J_{n,z}\left(\frac{2 \pi f r_z}{c}\right) e^{\jj n \phi_{l}} \sum_{p=0}^{P-1} \frac{\E^{\jj(m-n) \phi_{p,z}}}{P}.
\end{aligned}
\end{equation}

\noindent Although the latter expression may seem impractical from a computational perspective, note that the rightmost term $ (1/P)\sum_{p=0}^{P-1} \E^{j(m-n) \phi_{p,z}}$ becomes zero when $n \neq m$. Otherwise, when $n = m$ this term is one, leading to:
\begin{equation}
\widehat{H}_{m, l, z}(f) = H_{l,z}(f)\, \mathrm{j}^m J_{m,z}\left(\frac{2 \pi f r_z}{c}\right) \E^{\jj m \phi_{l}}.
\end{equation}

\noindent Thus, this expression contains the phase- and frequency-decoupled terms, $\E^{\jj m \phi_{l}}$ and $\mathrm{e}^{\mathrm{j} 2 \pi f \tau_{l,z}}$, the latter implicitly included in $H_{l,z}(f)$. Additionally, this excited frequency response $\widehat{H}_{m, l, z}(f)$ includes the frequency-dependent component, which can be eliminated by the optimal choice of an inverse filter $W_{m,z}(f)$. Mathematically, the phase-mode response $H_{m, l, z}(f)$ can be calculated as
\begin{equation} \label{simp_Hmlz}
\begin{aligned}
&\mkern-102mu H_{m, l, z}(f) = \widehat{H}_{m, l, z}(f)\, W_{m,z}(f)\\
&\mkern-102mu = H_{l,z}(f)\, \mathrm{j}^m J_{m,z}\left(\frac{2 \pi f r_z}{c}\right) W_{m,z}(f)\, \E^{\jj m \phi_{l}} \\
&\mkern-102mu =H_{l,z}(f)\, \E^{\jj m \phi_{l}} = \kappa_l \, \mathrm{e}^{\mathrm{j} 2 \pi f \tau_{l,z}} \,\mathrm{e}^{\jj m \phi_{l}},
\end{aligned}
\end{equation}

\noindent with
\begin{equation} \label{filtro_W}
W_{m, z}(f)=\frac{2}{\mathrm{j}^m\left[J_{m, z}\left(\frac{2 \pi f r_z}{c}\right)-\mathrm{j}\, J_{m, z}^{\prime}\left(\frac{2 \pi f r_z}{c}\right)\right]},
\end{equation}

\noindent where $J_{m, z}^{\prime}\left(\cdot\right)$ is the first derivative of $J_{m, z}\left(\cdot\right)$. This filter has demonstrated to be an optimal choice since $J_{m, z}^{\prime}\left(\cdot\right)$ avoid deep nulls in $W_{m, z}(f)$, providing larger bandwidth for the estimation and performing an accurate estimation of the azimuth angle $\phi_l$ for several angles $\theta_l$ \cite{Zhang_2017}.

In order to improve the estimation, the phase-mode average domain between all the circles at different heights $z$ is calculated, being the phase-mode expansion
\begin{equation} \label{suma_concentrica}
H_{m, l}(f)=\frac{1}{P} \sum_{z=z_0}^{z_{P-1}} H_{m, l, z}(f) \approx H_{l}(f) e^{\mathrm{j} m \phi_{l}}.
\end{equation}

\noindent The distribution of the $P$ circles in the XY plane is shown in Fig. \ref{esquema_cortes}. These circles are distributed in the range of heights $z_i \in [-\rho,\rho]$. Previous work has shown that the use of multiple arrays significantly improves the estimation of DoA and ToA \cite{phase_mode2007, AlexElipses2023, AlexEuCAP2023}. $H_{l}(f)$ stands for the frequency response at the origin. Additionally, note that weights $w_i$ with $\sum_{i}w_{i} = 1$ might be included in \eqref{suma_concentrica} to prioritize a specific subset of circles within the toric array in the estimation process. Although the use of weights is not mandatory in this development, as the toric geometry ensures a minimum radius $r_z$, these weights might be necessary in other three-dimensional geometries where one of the circle estimations degrades the overall characterization.

In multipath environments ($l > 1$), the phase-mode response for every wave impinging the toric array is the sum of the individual contribution from each wave, which leads to
\begin{equation} \label{suma_Hm}
H_{m}(f) = \sum_{l} H_{m, l}(f) = \sum_{l}  H_{l}(f)\, \E^{\mathrm{j} m \phi_{l}}.
\end{equation}

\noindent The 2-D Discrete Fourier Transform (DFT) of the former expression provides the joint azimuth DoA and ToA estimation as:

\begin{equation}
\widetilde{\mathbf{H}}(\phi, \tau) = \sum_{m} \sum_{k} H_{m}(f) \, \E^{- \jj   \left(\frac{m \phi}{M}+\frac{2 \pi (f_\textrm{min} + k f_s ) \tau}{K}\right)},
\end{equation}

\noindent where $K$ stands for the number of frequency samples and frequency spacing $f_s = B/K$ with $B$ being the signal bandwidth and $M$ is the total number of considered phase modes. The term $\,f_{min}$~is the lowest frequency considered in the bandwidth and $m$ and $k$ are the indexes that characterize M phase modes and K frequency samples, respectively. The ToA $\tau_l$ is extracted through the frequency $f$ obtained in $\mathrm{e}^{\mathrm{j} 2 \pi f \tau_{l}}$, while $\phi_l$ is related to the phase-mode domain in the complex exponential $e^{\mathrm{j} m \phi_{l}}$ [see \eqref{simp_Hmlz}]. Consequently, the time domain resolution and the maximum observable time are given by $1/B$ and $(K\nolinebreak-\nolinebreak1)/B$. Similarly, the angular resolution is given by $2\pi/M$ in the phase domain. The previous expression can be efficiently calculated through the Fast Fourier Transform (FFT) algorithm. Note that, as a 2D-DFT based method, the resolution in the angular and time domains is determined by $B$ and $M$. Therefore, it is advised to consider ultra-wideband channels and a high number of phase modes to maximize the resolution, and thereby mitigate the off-grid phenomena.

\subsection{Elevation of Arrival (EoA) and Time of Arrival (ToA)}

The second step takes advantage of the previous estimation of the azimuth $\phi_l$ to accurately estimate the elevation angle $\theta_l$ and the time of arrival $\tau_l$. Basically, the same $P^2$ spatial samples previously defined to generate $P$ circles with $P$ samples in the XY plane can be used to form circles in the planes defined by the $\phi-Z$ axes. These $P$ circles, also with $P$ samples per circle, are those contained in perpendicular cuts to the tube [see Fig. \ref{esquema_toros}(b)].

Similarly to (\ref{H_lz}), the frequency response at the center of the tube, given an angle $\phi_l$, is

\begin{equation}
H_{l,\phi_l}(f)=\eta_l \, \mathrm{e}^{\mathrm{j} 2 \pi f \tau_{l,\phi_l}},
\end{equation}

\noindent where $\eta_l$ is the complex attenuation, and $\tau_{l,\phi_l}$ is the delay of the wave $l$ from the source to the tube center at $\phi_l$. Therefore, the frequency response at the $p$-th sample reads 
\begin{equation} \label{Hplz_v3}
H_{p, l, \phi_l}(f)=\left(\frac{d_{l,\phi_l}}{d_{p, l, \phi_l}}\right)^{\gamma/2} H_{l,\phi_l}(f)\, \E^{\jj 2 \pi f \Delta d_{p, l, \phi_l} / c}\, ,
\end{equation}

\noindent with

\begin{equation}
\Delta d_{p,l,\phi_l} = d_{l,\phi_l} - \sqrt{d_{l,\phi_l}^2+\rho^2-2 d_{l,\phi_l} \rho \cos \left(\theta_{l}-\theta_{p,\phi_l}\right)}.
\end{equation}

\noindent After Taylor's series expansion (\ref{taylor_series}), we arrive to

\begin{equation} \label{Hplz_v4}
H_{p, l, \phi_l}(f)=\left(\frac{d_{l,\phi_l}}{d_{p, l, \phi_l}}\right)^{\gamma/2} H_{l,\phi_l}(f)\, \E^{\jj 2 \pi f  \rho \cos \left(\theta_{l}-\theta_{p,\phi_l}\right) / c}\, ,
\end{equation}

\noindent Note that, by choosing the ring coincident with the $\phi_l$ estimation, we ensure to align the wave incidence plane with the distribution of the $P$ tube samples. Thus, the sine dependent variable term that appeared in eqs. (\ref{delta_dplz_v2}), (\ref{taylor_series}) and (\ref{Hplz_v2}), becomes constant and is removed from eq. (\ref{Hplz_v4}). Visually, the chosen circular array laying in the $\Phi_l$Z plane is illustrated in Fig. \ref{esquema_cortes}. This fact ensures the correct estimation of $\theta_l$ and $\tau_{l,\phi_l}$ as the optimal estimation is provided when the plane of incidence coincides with the plane where the spatial sampling lies \cite{AlexElipses2023}.

Therefore, considering a negligible attenuation between the edge and the center of the tube, $H_{p, l, \phi_l}(f)$ is given by

\begin{equation} \label{Hplz_v5}
H_{p, l, \phi_l}(f)= H_{l,\phi_l}(f)\, \E^{\jj 2 \pi f  \rho \cos \left(\theta_{l}-\theta_{p,\phi_l}\right) / c}\, ,
\end{equation}

The former expression is similar to the one proposed in (\ref{Hplz_simplificada}), with the difference of: (i) considering a radius $\rho$; (ii) the cosine term depends on the elevation $\theta_l$; and (iii) the considered center is that of the tube and not that of the torus. Hence, a development in the form of basis functions $\E^{\jj m\theta_{p,\phi_l}}$, and the choice of an optimal $W_{m, \phi_l}(f)$ filter, similar to that in (\ref{JacobiAnger_ecuacion})-(\ref{filtro_W}), results in 
\begin{equation} \label{suma_mlphi}
H_{m, l, \phi_l}(f) = \eta_l \, \mathrm{e}^{\mathrm{j} 2 \pi f \tau_{l,\phi_l}} \mathrm{e}^{\jj m \theta_{l}}.
\end{equation}

\noindent The summation of the multiple incident waves (\ref{suma_Hm}) leads to $H_{m, \phi_l}(f)$. Finally, the application of the 2-D FFT to $H_{m, \phi_l}(f)$ provides the joint angular-time domain for the DoA elevation angle $\theta$ and the ToA at the center of the tube $\tau_{\phi_l}$, i.e., $\widetilde{\mathbf{H}}(\theta, \tau_{\phi_l})$. According to the torus geometry, the delay at the center of the tube, $\tau_{l,\phi_l}$, is related to the delay at the center of the torus, $\tau_{l}$, as follows:
\begin{equation} \label{tau_centro_a_tubo}
\tau_{l} - \tau_{l,\phi_l} = \frac{R \,  \lvert \sin (\theta_l) \rvert}{c} .
\end{equation}

\noindent This last step completely characterizes the DoA and ToA of the $l$-th wave by extracting the angles $\phi_l$, $\theta_l$, and the time $\tau_l$.

\subsection{Considerations of the method}

Subsections II.A and II.B have presented the theoretical framework for the 3D characterization of the communication channel. Although the method has been shown to work correctly in 2D scenarios with several geometries, due to some simplifications that are carried out, it is necessary to perform a correct assignment of the considered parameters. These approximations involve the appearance of artifacts, i.e. spectral contributions in $\widetilde{\mathbf{H}}(\phi, \tau)$ and $\widetilde{\mathbf{H}}(\theta, \tau_{\phi_l})$ that may mislead with the real incident wave. As long as these artifacts are controlled, the characterization can be properly performed. For this purpose, we define a metric $\Delta$, which indicates the difference between the power of the incident signal and the largest artifacts. Therefore, $\Delta$ shows the array response of the toric geometry when a wave with certain parameters $(\phi_l,\theta_l,\tau_l)$ impinges on the array given the joint angular-time domains $\widetilde{\mathbf{H}}(\phi, \tau)$ and $\widetilde{\mathbf{H}}(\theta, \tau_{\phi_l})$. If the value of $\Delta$ is greater than 0~dB, the incident signal is distinguishable from the artifacts [\citenum{AlexElipses2023},~Fig.~3]. Thus, large $\Delta$ values maximize the dynamic range when characterizing the communication channel.

Concerning the number of spatial samples $P$, it must be chosen in such a way that we ensure compliance with the Nyquist spatial sampling theorem, i.e., a separation between samples less than half wavelength. Otherwise, spatial aliasing appears, which translates into an increase in the level of the artifacts. Fundamentally, this depends on the radii considered for the circles in the different planes of the torus. In Section II.A, the boundary is given by the outer circle when $z = 0$, where $r_z = R + \rho$, while in Section II.B, the considered radius is directly $\rho$.

Given the maximum value of $r_{z}$, we can ensure that the sampling theorem is satisfied on the torus with $P$ equispaced samples per circle if
\begin{equation} \label{Nyquist}
2  (R+\rho)  \sin(\pi/P) < \lambda/2.
\end{equation}
Regarding the number of phase modes $M$, it can be demonstrated that $J_n(\cdot) \approx 0$ for a sufficiently large order $n$ \cite{phase_mode2008, Jacobi-Anger}. Thus, $M$ must be chosen so that it is lower than the above threshold. Otherwise, the denominator in \eqref{filtro_W} would tend to zero, generating numerical instabilities in the method. 

 Logically, the greater the number of filters, the greater the required computational time. The number of raw $W_{m,z}(f)$ filters needed is $M\times P$ in the first step. However, by taking advantage of the reflection (mirror) symmetry of the torus on the $z$-axis, the number of filters can be decreased to $M \times (P/2 +1)$. Also, by taking advantage of the symmetry of the Bessel functions for positive and negative indexes $m$ [$J_{-m}(x) = (-1)^m J_{m}(x)$], the total number of filters is reduced to $(M/2 + 1) \times (P/2 +1)$. In the second step, only a single circle is considered, thus,  $M/2 + 1$ filters are needed in this case. In total,  $(M/2 + 1) \times (P/2 +2)$ filters are required for a joint characterization of the AoA, EoA and ToA.

\begin{figure}[t]
	\centering
	\includegraphics[width= 1\columnwidth]{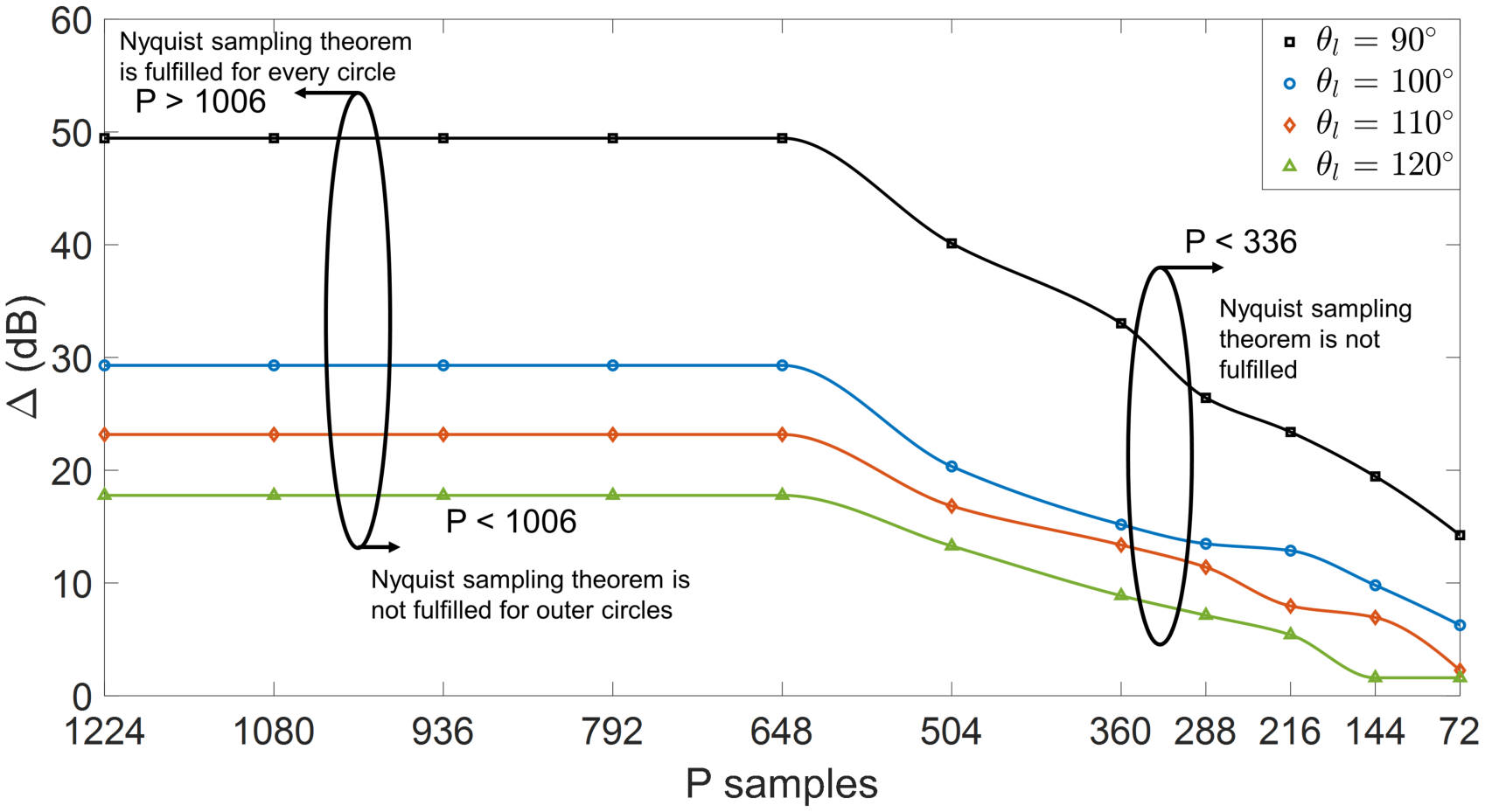}
	\caption{Relation between the power of the estimated wave and artifacts ($\Delta$) as a function of the number of spatial samples $P$ for several incident elevation angles $\theta_l$. Parameters of the considered scenario: $R = 0.5$~m, $\rho = 0.25$~m, $\phi_l = 180\degree$ and $\tau_l = 15$~ns. Non-compliance with the spatial sampling theorem increases the artifacts, thus decreasing the value of $\Delta$ in the estimation.} 
	\label{barrido_P}
\end{figure}

\section{\label{Results_theory} Validation of the method}

This Section discusses  and validates several aspects of the method presented in Section II. Specifically, a parametric study shows the main implications of the different parameters involved in the joint characterization of the AoA, EoA and ToA. According to the torus geometry, several aspect ratios, i.e., $R/\rho$, and sizes are considered to show the good performance of the method for diverse cases. Later, the joint estimation is validated through simulations at several frequency ranges for different scenarios based on both single-path and multipath scenarios.

\subsection{Parametric analysis}

Regarding the physical geometry of the torus, $R$, $\rho$ and $P$ are the main parameters to be considered. As previously stated, these are directly related to the Nyquist spatial sampling theorem [see \eqref{Nyquist}]. In order to analyze the effect of fulfilling this theorem, it is considered a case with a single-path scenario impinging the toric array with $\tau_l = 15$ ns ($d_l = 4.5$ m), and $\phi_l = 180\degree$ for several values of $\theta_l$. The torus size is given by $R = 0.5$ m and $\rho = 0.25$ m, with a frequency range that goes from 28 GHz to 32 GHz ($B = 4$ GHz) and $K = 200$~frequency samples. $M = 300$~phase modes are considered.  Given $R$, $\rho$, and $f_{max} = 32$~GHz ($\lambda_{min} = 9.37$ mm), the Nyquist theorem is satisfied even for outer circles ($r_z = 0.75$~m) if $ P > 1006$. For $336 < P < 1006$, it is exclusively satisfied for some inner circles. If $ P < 336$, it is not fulfilled even for the inner circles ($r_z = 0.25$~m).

For the given parameters, Fig. \ref{barrido_P} illustrates the value of the metric $\Delta$ when the number of samples $P$ is varied. Note that a higher value of $\Delta$ implies a better DoA and ToA estimation, as the level of the artifacts is reduced. For the case $\theta_l = 90\degree$, the  quality of the estimation is optimal due to the approximation performed in \eqref{Hplz_simplificada}. However, even non-coincident waves in the elevation plane ($\theta_l \neq 90\degree$) provide good results for $\Delta$. 

Concerning the number of spatial samples $P$, a flat behavior is observed in Fig. \ref{barrido_P} when the Nyquist theorem is fulfilled ($P>1006$). This implies that the estimation will not improve even if the number of samples is increased. In an intermediate region ($336<P \leq 1006)$, even though sampling is not performed correctly for the outer circles, a flat behavior is still observed until the sampling theorem is no longer satisfied for about half of the circles, i.e., $r_z = R$. This fact contrasts with estimation on single arrays, where not complying with the sampling theorem dramatically increases the artifacts \cite{AlexElipses2023}. The geometry of the torus, which supports estimation from concentric circles [see \eqref{suma_concentrica}], allows the number of $P$ sensors to be reduced below the sampling theorem, compensating for this decrease with the larger number of arrays deployed. When $P$ is further reduced ($P<336$), the value of $\Delta$ begins to significantly decay. In any case, note that despite the presence of artifacts, $\Delta > 0$~dB, which makes the incident wave distinguishable from the artifacts even for a reduced number of samples.

\begin{figure}[t]
	\centering
	\includegraphics[width= 1\columnwidth]{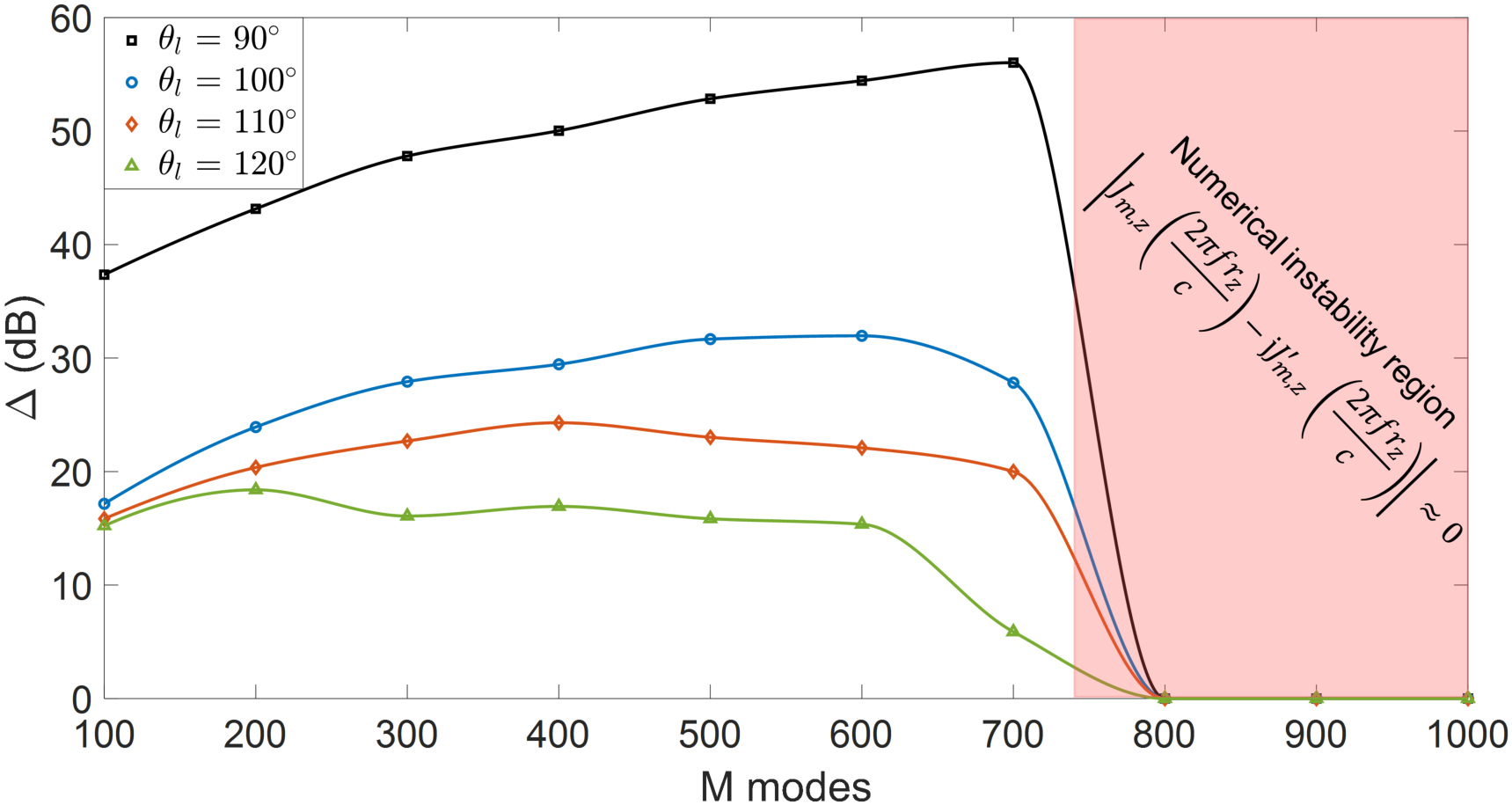}
	\caption{Relation between the power of the estimated wave and artifacts ($\Delta$) as a function of the number phase modes $M$ for several elevation angles $\theta_l$. Parameters of the considered scenario: $R = 0.75$~m, $\rho = 0.25$~m, $\phi_l = 180\degree$ and $\tau_l = 15$~ns.} 
	\label{barrido_M}
\end{figure}

In a second study, the influence of the number of phase modes $M$ in the DoA and ToA estimation is analyzed. All the parameters remain the same as in the previous experiment, except from the torus geometry. The parameters of the torus are: $R = 0.75$~m, $\rho = 0.25$~m and $P = 1440$. Fig. \ref{barrido_M} shows the value of $\Delta$ when varying the number of considered phase modes $M$.  Above a certain threshold, approximately $M > 700$ in this case, $J_n(\cdot) \approx 0$. This causes numerical instabilities in the computation of the filter $W_{m, z}(f)$ [eq. \eqref{filtro_W}] as the denominator approaches zero. As a result, the quality of the estimation significantly degrades. In the range $300 < M < 600$, an almost flat region is found, showing that, regardless of the number $M$ chosen, the estimation performs correctly. Finally, if $M$ is even more decreased, the estimation tends to degrade, besides decreasing the angular resolution (see Sec. II.A). Ideally, one should operate in the nearly flat region where $\Delta$ approaches the maximum. This region depends exclusively on the $2 \pi f r_z / c$ argument of the Bessel function, so a prior analysis of the optimal working regions is essential.

\begin{figure}[!t]
    \centering
    \subfigure[]{\includegraphics[width=1\columnwidth]{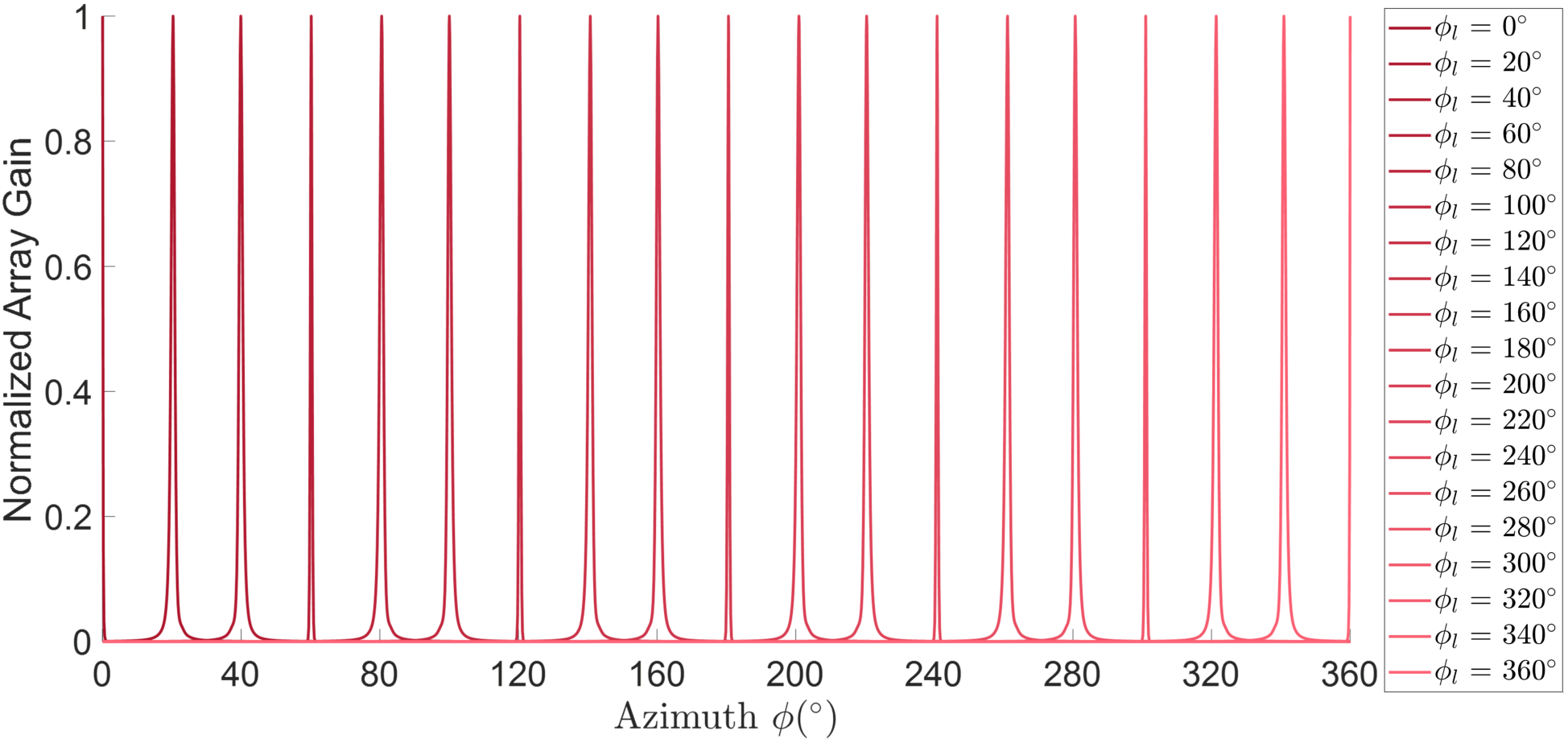}
	} 
    \subfigure[]{\includegraphics[width= 1\columnwidth]{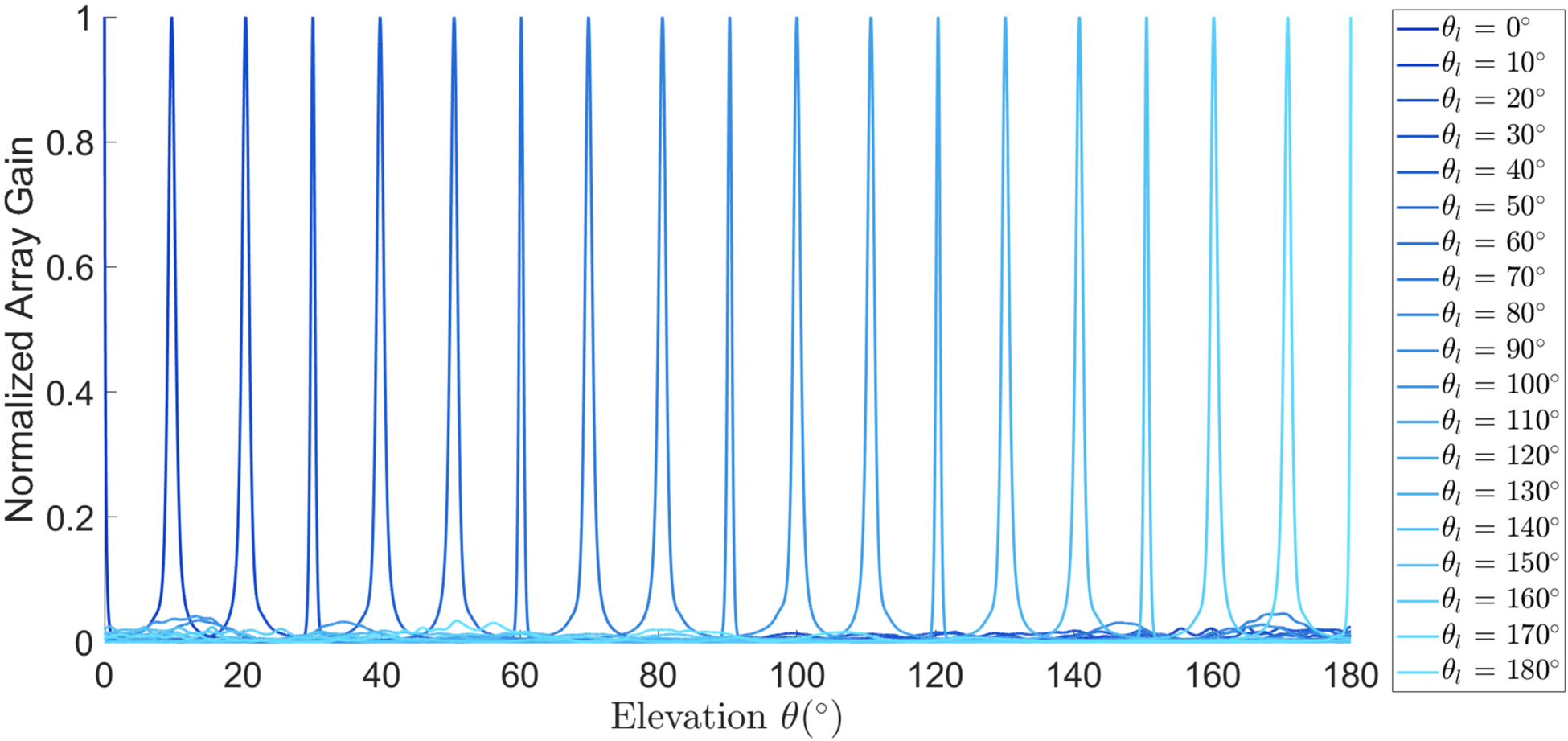}
	}
    \subfigure[]{\includegraphics[width= 1\columnwidth]{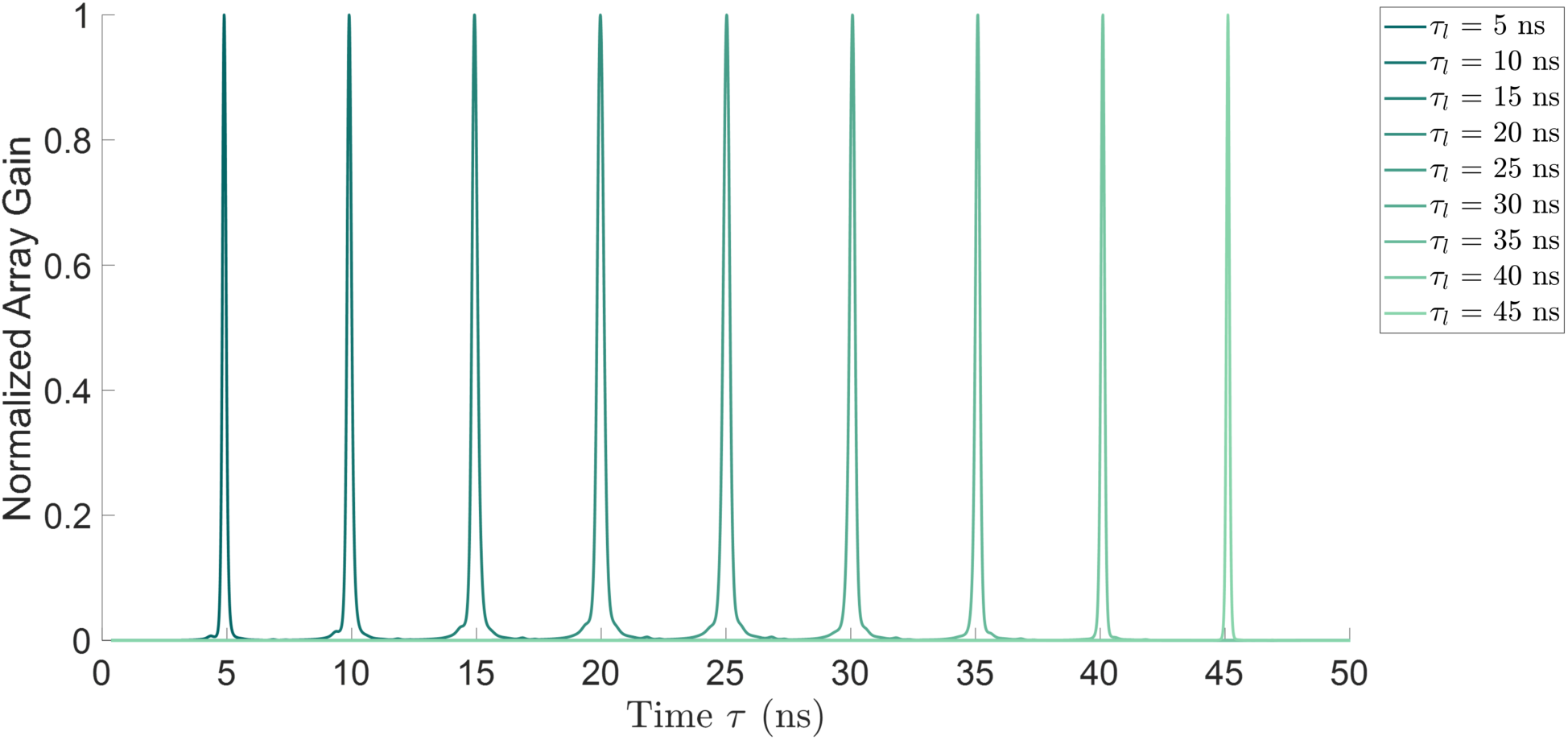}
	}
 
	\caption{Normalized array gain given a toric array with $R = 0.5$~m, $\rho~=~0.25$~m, $P = 720$, $K = 200$, $M = 300$ and $f \in [28,32]$~GHz for azimuth, elevation and time domains. (a) Azimuth (AoA) domain for $\phi_l \in [0\degree, 360\degree]$ with $20\degree$ steps $(\theta_l = 90\degree, \tau_l = 25 \textrm{ ns})$. (b) Elevation (EoA) domain for $\theta_l \in [0\degree, 180\degree]$ with $10\degree$ steps $(\phi_l = 180\degree, \tau_l = 25 \textrm{ ns})$. (c) Time (ToA) domain for $\tau_l \in (0 \textrm{ ns}, 50\textrm{ ns})$ with $5 \textrm{ ns}$ steps $(\phi_l = 180\degree, \theta_l = 90\degree)$.} 
	\label{array_gain}
\end{figure}

\begin{figure}[t]
	\centering
	\includegraphics[width= 1\columnwidth]{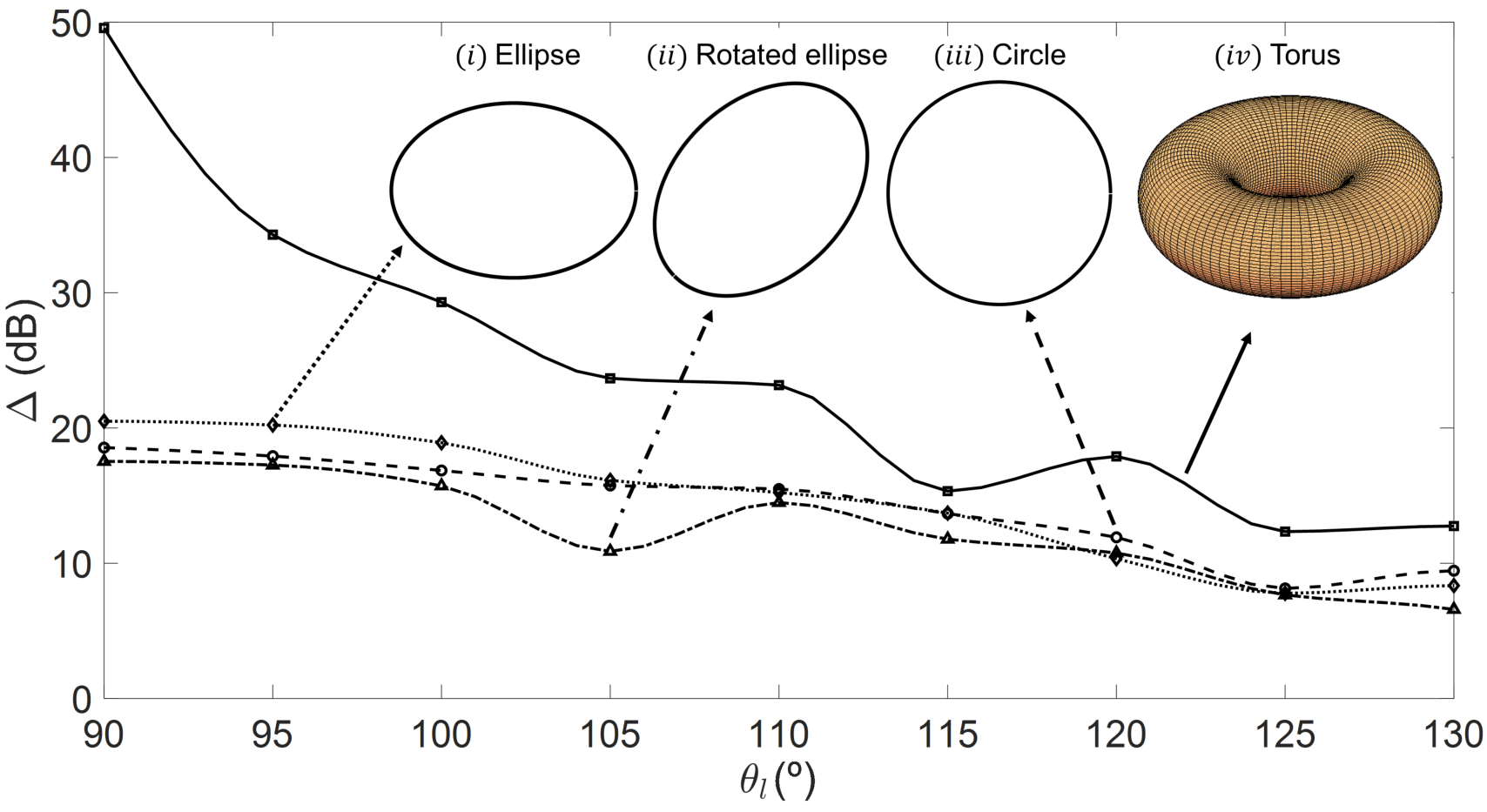}
	\caption{Degradation of the estimation as $\theta_l$ increases for the joint AoA ($\phi_l = 180\degree$) and ToA ($\tau_l = 15$~ns) values. Four different geometries are considered: (i) elliptical array, (ii) rotated elliptical array, (iii) circular array, and (iv) toric array.} 
	\label{barrido_theta}
\end{figure}

From a single AoA/EoA/ToA domain perspective, the array gain of the toric geometry can be obtained by averaging a single dimension of the phase-mode $H_{m,l}(f)$ [see \eqref{suma_concentrica}] or $H_{m,l,\phi_l}(f)$ [see \eqref{suma_mlphi}], given the pairs \{$\phi_l$, $\tau_l$\} or \{$\theta_l$, $\tau_{l,\phi_l}$\}, respectively. Fig. \ref{array_gain} shows the normalized array gain for several $\phi_l$, $\theta_l$ and $\tau_l$ values given the AoA/EoA/ToA domains. The high directivity, whether in angular or temporal domain, is illustrated across the three domains, which will enable accurate simultaneous characterization of azimuth, elevation and time of arrival. Finally, to demonstrate the potential of the toric geometry, the metric $\Delta$ is compared with previous two-dimensional geometrical approaches that make use of frequency-independent beamformers and phase-mode transformations. For this purpose, we consider the same incident wave ($\phi_l$ and $\tau_l$) as in the previous experiments, and the same frequency range, B and K. Fig. \ref{barrido_theta} shows the value of $\Delta$, given the azimuth (AoA) -- time (ToA) domain, for different elevation angles $\theta_l$ given four different geometries: (i) elliptical array, (ii) rotated elliptical array, (iii) circular array and (iv) toric array. Therefore, Fig. \ref{barrido_theta} compares the performance of FIBs combined with phase-mode transformations for several arrangements. Specifically, the theoretical framework for elliptical arrangements has been developed in our previous works \cite{AlexElipses2023, AlexEuCAP2023}, while the framework for circular arrangements is based on \cite{phase_mode2007, Zhang_2017}. The number of phase modes is fixed to $M=300$. Both elliptical arrays have a semi-major axis of 0.5 m, an eccentricity of 0.7, and $P=720$. Additionally, the rotated ellipse includes an angle rotation of $45\degree$. The single circular array has a radius of 0.5 m, and $P=720$. The toric array parameters are $R=0.5$~m, $\rho = 0.25$~m and $P = 720$. It can be appreciated that the toric array estimation outperforms all other geometries. This is due to the joint estimation based on the use of the concentric circular arrays that are discerned in the toroid geometry itself. Note that this sample distribution will later allow us to perform the accurate estimation of the elevation angle $\theta_l$, while in the other geometries, it is not possible due to the two-dimensional distribution of the samples. The application of the two-stage method might imply the presence of propagation errors if $\theta_l$ is calculated based on an inaccurate characterization of $\phi_l$, or vice versa when $\phi_l$ is initially calculated with an unknown $\theta_l$. Nevertheless, Fig. 6 illustrates the robustness of the method given the characterization for several angles.

In summary, the proposed method is shown to improve the quality of the estimation by decreasing the level of the artifacts compared to other geometries, such as single circles or ellipses, even enabling the estimation with under-sampling conditions due to the use of several arrays, i.e., several FIBs, simultaneously.

\subsection{Single-path Characterization}

\begin{figure}[!t]
	\centering
	\subfigure[]{\includegraphics[width=1\columnwidth]{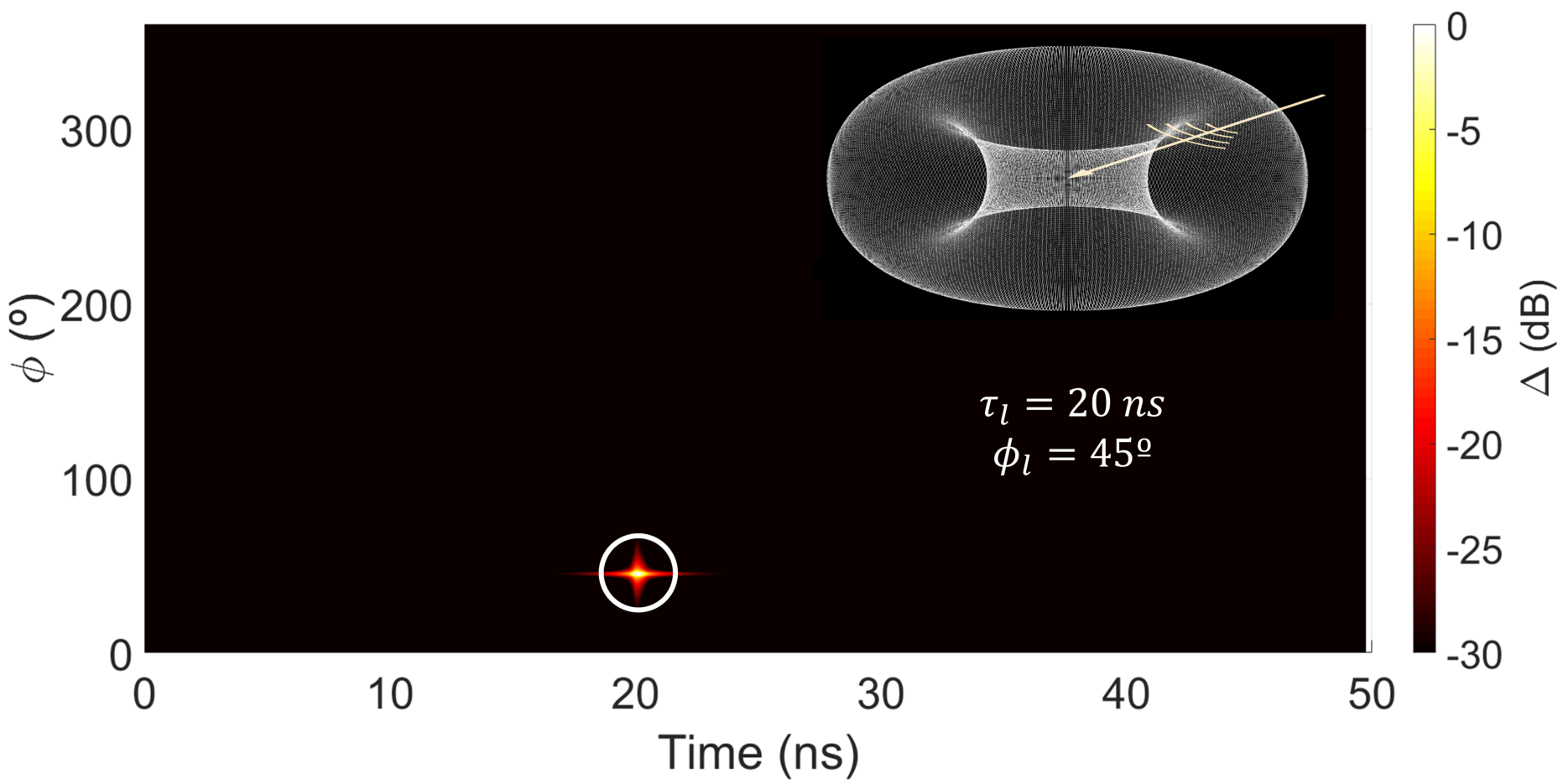}
	} 
\subfigure[]{\includegraphics[width= 1\columnwidth]{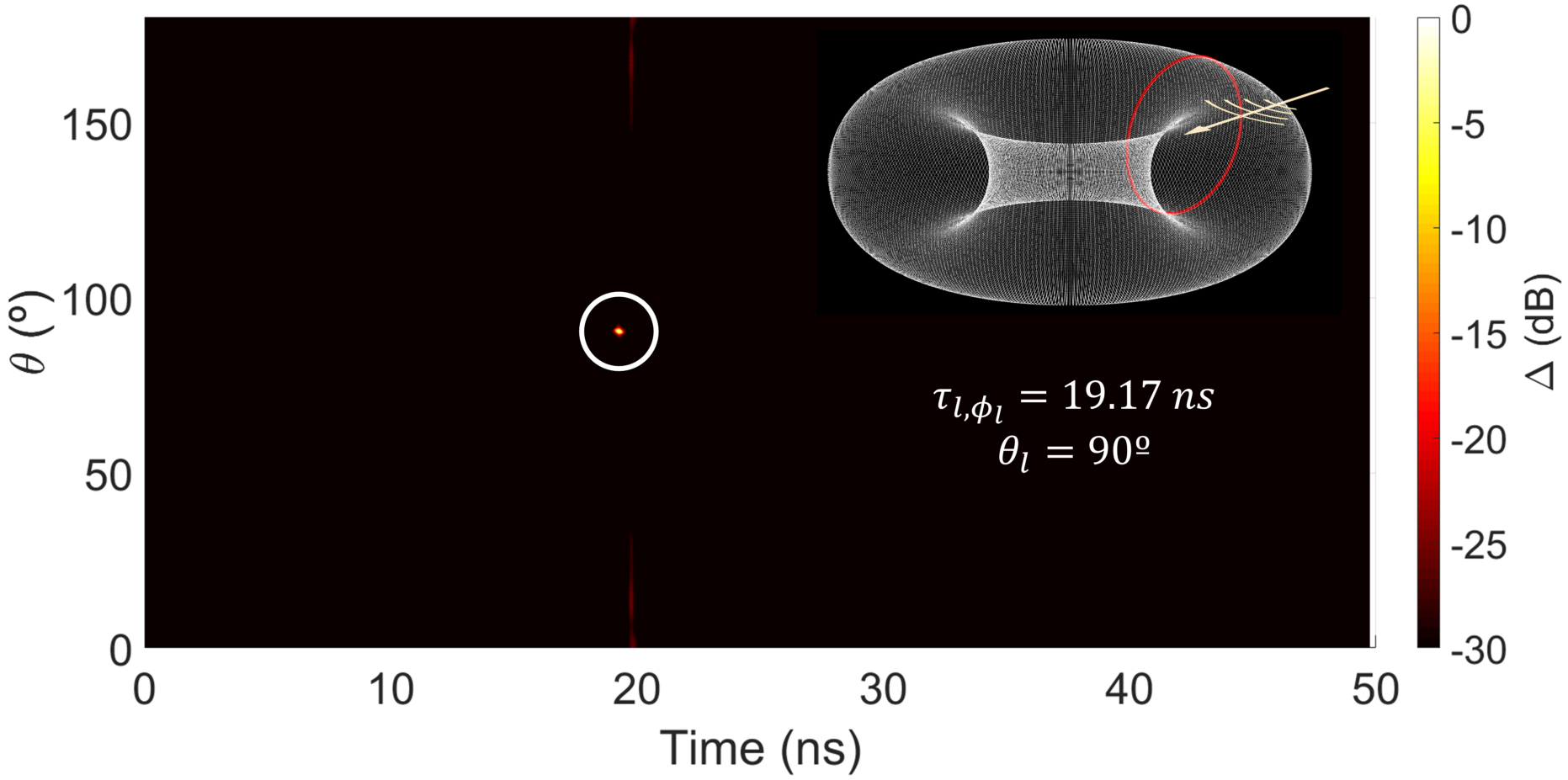}
	}
	\caption{Joint azimuth ($\phi_l = 45\degree$), elevation ($\theta_l = 90\degree$), and time ($\tau_l\nolinebreak=\nolinebreak20$~ns) of arrival estimation for a toric array. (a) Azimuth--time domain. (b) Elevation--time domain. Parameters of the scenario: $R = 0.25$~m, $\rho = 0.125$~m, $P = 720$, $M = 300$ and $f \in [58,62]$~GHz.} 
	\label{single_path_ej1}
\end{figure}

Once some of the key parameters for the method have been analyzed, this Section illustrates some examples of the 3D joint characterization of the AoA ($\phi_l$), EoA ($\theta_l$), and ToA ($\tau_l$). Let us assume 
 a frequency range $f \in [58,62]$~GHz with $B = 4$~GHz and $K = 200$ frequency samples. According to Section~II.A, for $R = 0.25$~m and $\rho = 0.125$~m, $P = 720$ spatial samples and $M = 300$ phase modes are enough to reduce the level of the artifacts and ensure a proper estimation. Let us assume a wave impinging the toric array with $\tau_l=\nolinebreak20$~ns, $\phi_l = 45\degree$, and $\theta_l = 90\degree$. Fig. \ref{single_path_ej1}(a) shows the azimuth--time domain extracted from the phase-mode transformation. As observed, the value of the metric $\Delta$ is maximum around the correct values of the AoA ($\phi_l = 45\degree$) and ToA ($\tau_l = 20$ ns). The level of the artifacts is below -30 dB, so they do not appear in the figure. This effect is obtained due to the joint estimation of the multiple concentric circles contained in the torus. 
 
 Given the azimuth estimation, the circumference lying on this angle $\phi_{l}$ is chosen. Fig. \ref{single_path_ej1}(b) illustrates the elevation--time domain. Note that the value of $\Delta$ is maximum at the EoA ($\theta_l = 90\degree$) and ToA ($\tau_{l,\phi_{l}} = 19.25$ ns). According to eq. \eqref{tau_centro_a_tubo}, the exact time of arrival should be $19.17$ ns, appearing this difference due to the temporal resolution $1/B$. Nonetheless, this value is within the resolution range of the method, being a correct $\tau_{l,\phi_{l}}$ estimation. In Fig. \ref{single_path_ej1}(b), the level of the artifacts is below -23 dB.

\begin{figure}[!t]
	\centering
	\subfigure[]{\includegraphics[width=1\columnwidth]{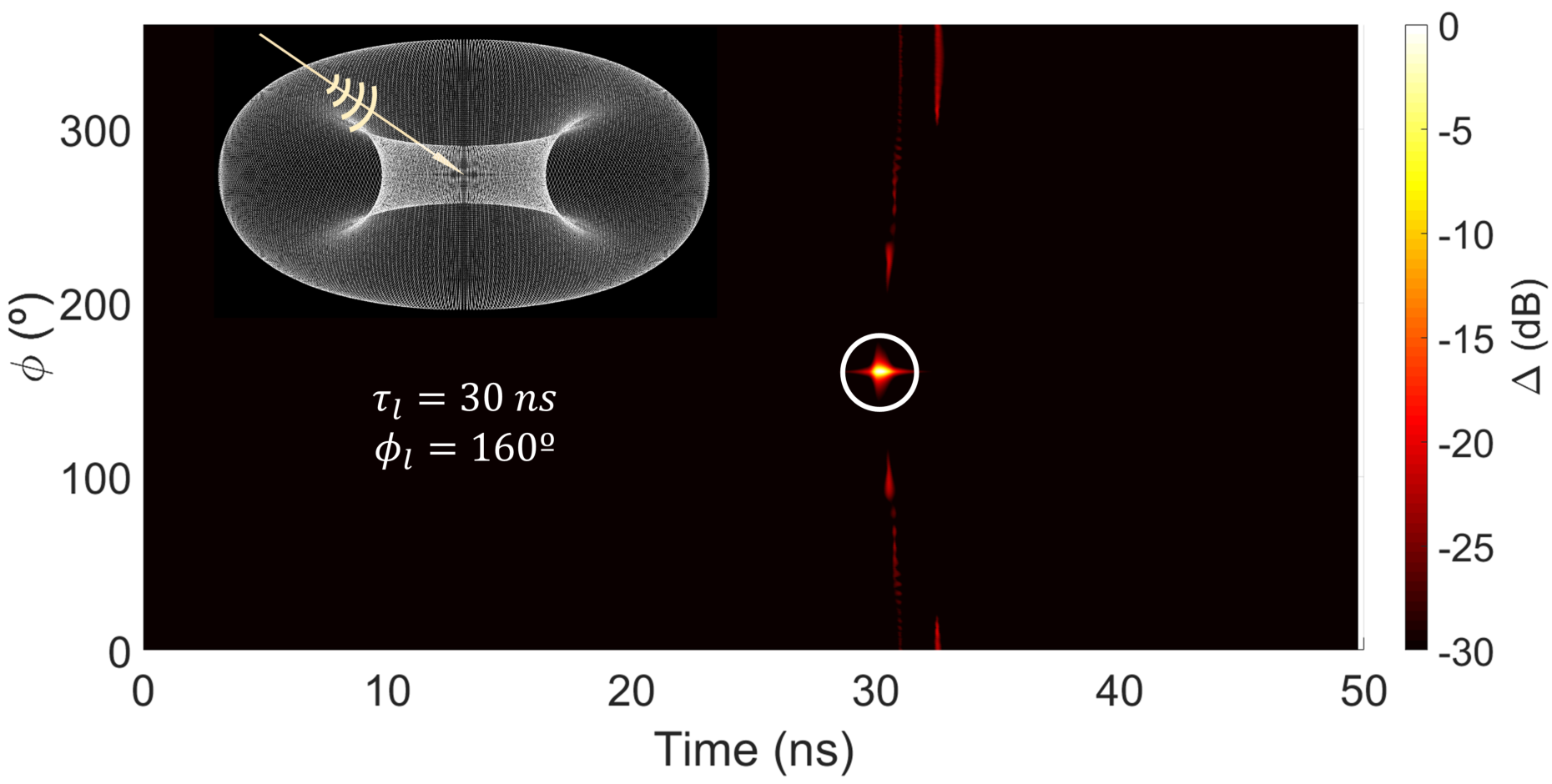}
	} 
\subfigure[]{\includegraphics[width= 1\columnwidth]{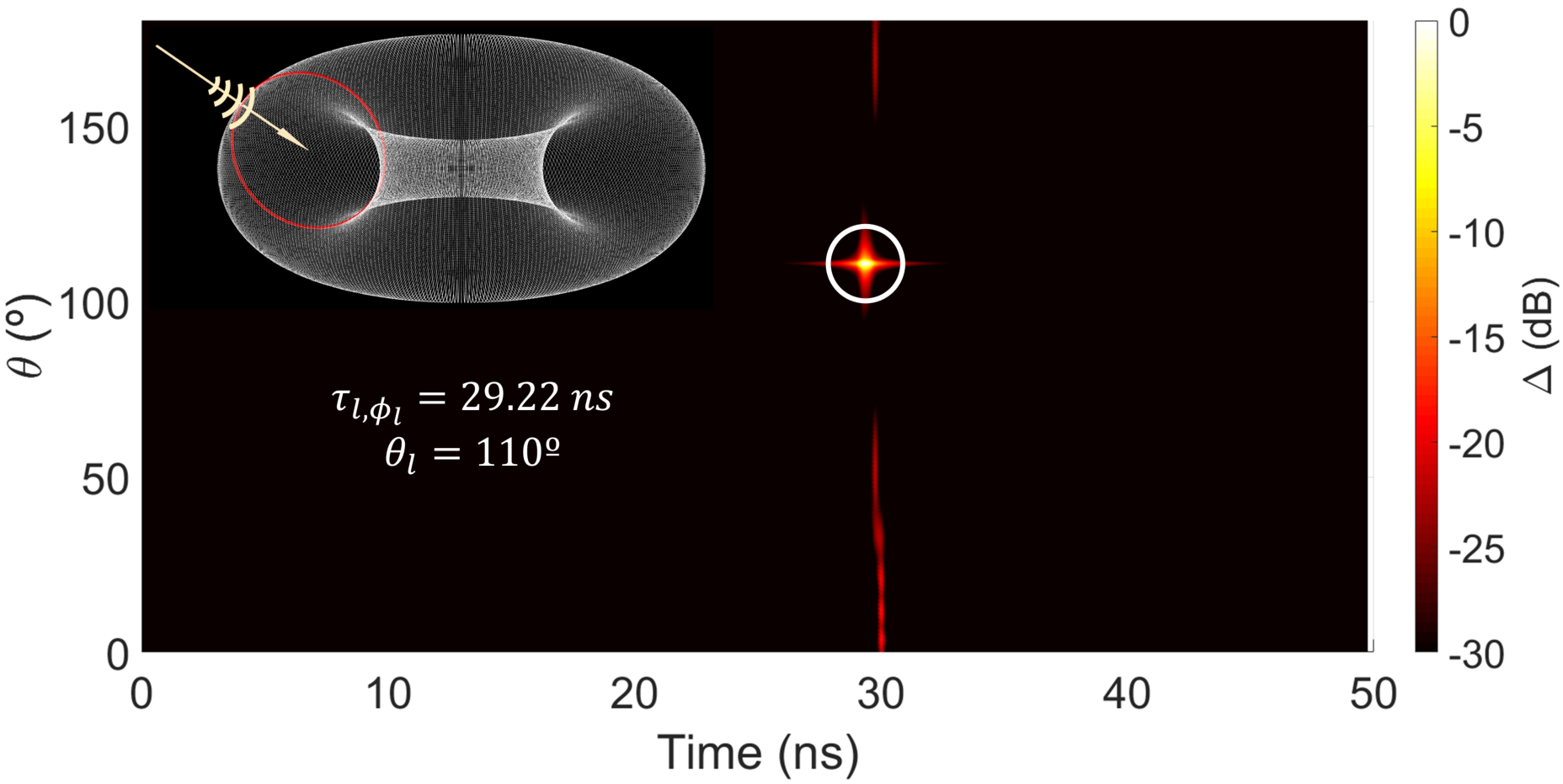}
	}
	\caption{Joint azimuth ($\phi_l = 160\degree$), elevation ($\theta_l = 110\degree$), and time ($\tau_l\nolinebreak=\nolinebreak20$~ns) of arrival estimation with a toric array. (a) Azimuth--time domain. (b) Elevation--time domain. Parameters of the scenario: $R = 0.25$~m, $\rho = 0.125$~m, $P = 720$, $M = 300$ and $f \in [58,62]$~GHz.} 
	\label{single_path_ej2}
\end{figure}

\begin{figure*}[!b]
	\centering
	\includegraphics[width= 0.97\textwidth]{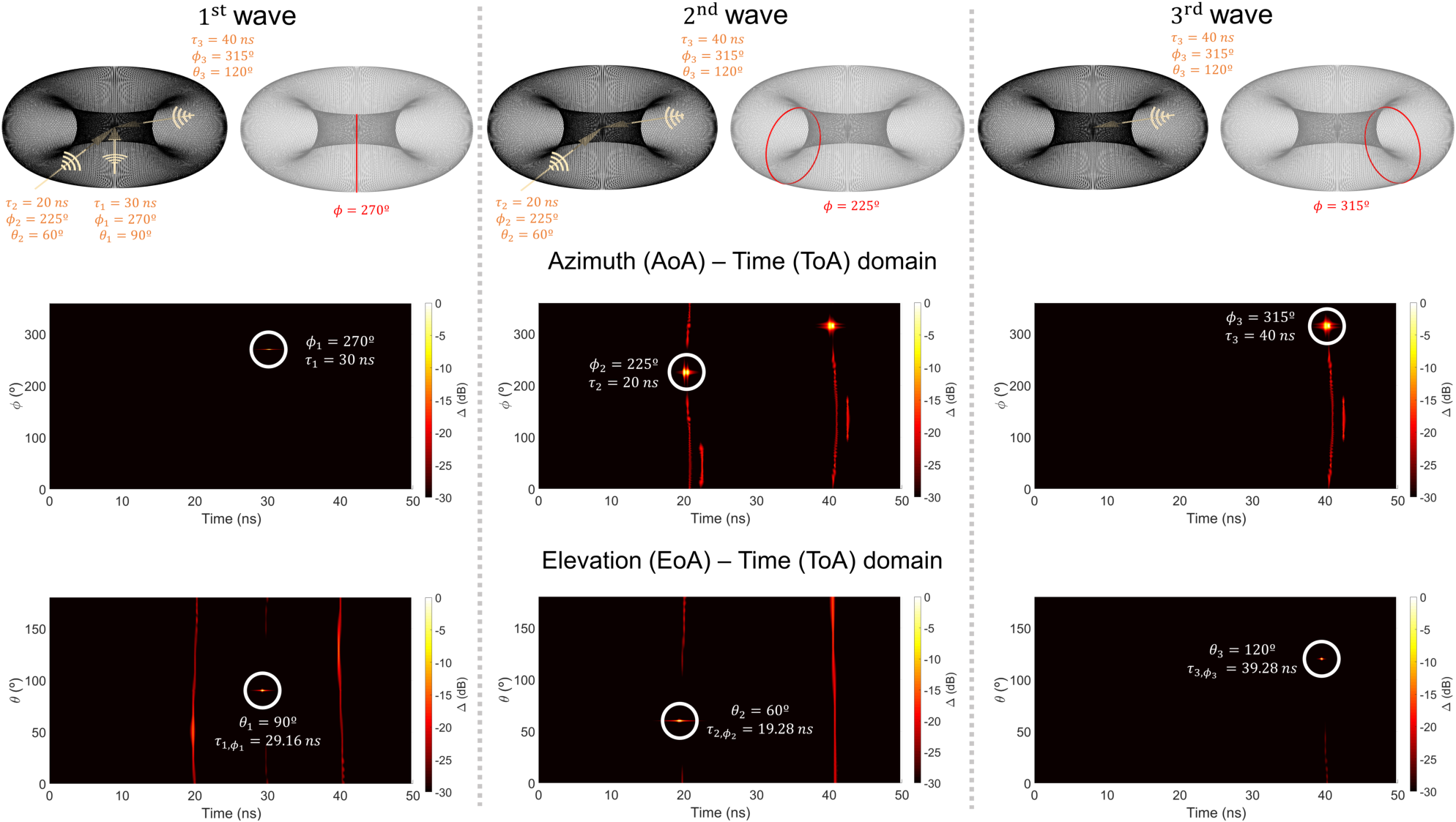}
	\caption{Joint azimuth, elevation, and time of arrival estimation for a toric array with $R = 0.25$~m, $\rho = 0.125$~m, $P = 720$, $M = 300$, $K = 200$ and $f \in [58,62]$~GHz. A multipath environment with $L = 3$ waves is considered. AoA: \{$\phi_1 = 270\degree$, $\phi_2 = 225\degree$, $\phi_3 = 315\degree$\}. EoA: \{$\theta_1 = 90\degree$, $\theta_2 = 60\degree$, $\theta_3 = 120\degree$\}. ToA: \{$\tau_1\nolinebreak=\nolinebreak30$~ns, $\tau_2\nolinebreak=\nolinebreak20$~ns, $\tau_3\nolinebreak=\nolinebreak40$~ns\}. Left, central and right panels show the estimation for the $\textrm{1}^{\textrm{st}}, \textrm{2}^{\textrm{nd}}$ and $\textrm{3}^{\textrm{rd}}$ wave, respectively.} 
	\label{estimacion_multipath}
\end{figure*}

 In a second experiment for the single-path scenario, we consider a wave with  $\phi_l = 160\degree$, $\theta_l = 110\degree$, and $\tau_l =30$~ns. The additional parameters remain identical to the previous experiment. Figs. \ref{single_path_ej2}(a) and \ref{single_path_ej2}(b) illustrate the azimuth--time and elevation--time domains, respectively, for the joint estimation of this incident wave. As observed, the maximum value of $\Delta$ coincides with the AoA, EoA and ToA positions in both figures. Artifacts amplitude appears 20.3 dB and 18.4 dB below the correct estimation in Figs. \ref{single_path_ej2}(a) and \ref{single_path_ej2}(b), respectively. Note that the temporal dispersion explained in Section II.A due to an elevation $\theta_l \neq 90 \degree$ is not appreciable because of the high working frequencies ($f_c = 60$~GHz), since the physical size of the array decreases notably, being $\tau_{l,z} \approx \tau_l$. Therefore, the estimation $\tau_l$ performed in the first step can be considered as reliable taking into account high frequencies in the \mbox{mm-wave} band with small array sizes where $\tau_l \gg 2\rho/c$. The two previous experiments have shown the ability to correctly identify waves in the three-dimensional space and in the mm-wave frequency range.

\subsection{Multipath Characterization}

The previous section has validated the performance of the FIB for the 3D channel characterization in a single-path scenario. According to \eqref{suma_Hm}, for a multipath scenario ($l > 1$), the spectral response of the beamformer is directly the sum of the responses of each incident wave. However, in practice, the amplitude of the spectral response is actually dependent on the elevation $\theta_l$ due to the assumption made in \eqref{Hplz_simplificada}. This fact was corroborated in Fig.~\ref{barrido_theta} with differences of up to 30~dB when going from $\theta_l = 90\degree$ to $\theta_l = 130\degree$. This implies that, under the assumption of a set of $L$ incident waves with similar $\theta_l$, the DoA-ToA domain provides a correct characterization of the scenario. However, if this set of waves has $\theta_l$ different from each other, the waves with $\theta_l$ close to $90\degree$ may fade the rest of the incident waves. Therefore, estimation in multipath environments with several $\theta_l$ values, i.e., the most general case, can be performed based on a subtraction strategy of the previously estimated waves. Given the estimation of the $l$-th path, the influence of the previous $l-1$ paths can be removed as $ \sum_{n = l}^{L}H_{p,l,z}(f) = \sum_{n = 1}^{L}H_{p,l,z}(f) - \sum_{n = 1}^{l-1}H_{p,l,z}(f)$. If the estimation is performed correctly, the effect of the substracted $l-1$ waves is removed from $\widetilde{\mathbf{H}}(\phi, \tau)$ and $\widetilde{\mathbf{H}}(\theta, \tau_{\phi_l})$. In case that the subtraction is not effective due to the temporal ($1/B$ seconds) and angular ($2\pi/M$ radians) resolution values, one can proceed to reconstruct $H_{p,z}(f)$ as a discretized set of waves based on a low-complexity search with expectation-maximization algorithms such as Space Alternating Generalized Expectation-maximization (SAGE) \cite{SAGE, Fan_2019}. Thus, by employing this approach, the off-grid phenomena can be mitigated.

To summarize the rationale previously described, Algorithm~1 presents a pseudocode that illustrates the main steps to be followed in a general multipath environment. The method, originally based on DoA and ToA estimation for circular arrays, has been generalized to toric arrays where the geometry of the torus has been exploited to define circles in multiple planes. Through an appropriate choice of the parameters involved in the method, good performance of the characterization will be obtained.

\begin{table}[!t]
\renewcommand{\arraystretch}{1.2}
\resizebox{\columnwidth}{!}{

\begin{tabular}{l}
\hlinewd{1.4pt} \textbf{Algorithm 1} Joint Characterization of Azimuth, Elevation \\
and Time of Arrival based on Toric Arrays. Multipath case.  \\
\hlinewd{1.4pt} \textbf{Input:} Channel Frequency Response samples \\

\quad \quad \quad \quad \!\!\!\!\!\!\! at the toric array, i.e., $H_{p,z}(f)$.\\

\textbf{Output:} Azimuth, Elevation and Time of Arrival\\

\quad \quad \quad \quad \{$\phi_l$, $\theta_l$, $\tau_l$\}. \\

{\footnotesize $\mathbf{1}$} \textbf{for} $l = 1:L$ \\

{\footnotesize $\mathbf{2}$} \quad \textit{Azimuth of Arrival (AoA)}  \\

{\footnotesize $\mathbf{3}$} \quad \quad \textbf{if} $l > 1$ \\

{\footnotesize $\mathbf{4}$} \quad \quad \quad \quad $H_{p,z}(f) = \sum_{n = l}^{L}H_{p,l,z}(f) $\\

{\footnotesize $\mathbf{5}$} \quad \quad \textbf{end} \\

{\footnotesize $\mathbf{6}$} \quad \quad $\widehat{H}_{m, z}(f)
= 1/P \sum_{p=0}^{P-1} H_{p, z}(f)\, \E^{\jj m \phi_{p,z}}$ \\ 

{\footnotesize $\mathbf{7}$} \quad \quad $H_{m, z}(f) = \widehat{H}_{m, z}(f)\, W_{m,z}(f)$  \\

{\footnotesize $\mathbf{8}$} \quad \quad $H_{m}(f) = \sum_{z} H_{m,z}(f)$ \\

{\footnotesize $\mathbf{9}$} \quad \quad $\widetilde{\mathbf{H}}(\phi, \tau) = \operatorname{2-D \,\, FFT}\{H_{m}(f)\}$ \\

{\footnotesize $\mathbf{10}$} \quad \quad $\phi_l$ = $\underset{\phi, \tau}{\mathrm{arg\,max}}\,\left(\widetilde{\mathbf{H}}(\phi, \tau)\right)$  \\

{\footnotesize $\mathbf{11}$} \quad \textit{Elevation of Arrival (EoA) and Time of Arrival (ToA)}  \\

{\footnotesize $\mathbf{12}$} \quad\, Choose $H_{p, \phi_l}(f)$ based on the estimated $\phi_l$\\

{\footnotesize $\mathbf{13}$} \quad\, $\widehat{H}_{m, \phi_l}(f) = 1/P \sum_{p=0}^{P-1} H_{p, \phi_l}(f)\, \E^{\jj m \theta_{p,\phi_l}}$ \\

{\footnotesize $\mathbf{14}$} \quad\, $H_{m, \phi_l}(f) = \widehat{H}_{m, \phi_l}(f)\, W_{m,\phi_l}(f)$ \\

{\footnotesize $\mathbf{15}$} \quad\, $\widetilde{\mathbf{H}}(\theta, \tau_{\phi_l}) = \operatorname{2-D \,\, FFT}\{H_{m,\phi_l}(f)\}$ \\

{\footnotesize $\mathbf{16}$} \quad\, $\{\theta_l,\tau_{l,\phi_l}\}$ = $\underset{\theta, \tau_{\phi_l}}{\mathrm{arg\,max}}\,\left(\widetilde{\mathbf{H}}(\theta, \tau_{\phi_l})\right)$ \\

{\footnotesize $\mathbf{17}$} \quad\, $\tau_{l} = R \,  \lvert \sin (\theta_l) \rvert / c + \tau_{l,\phi_l}$ \\

{\footnotesize $\mathbf{18}$} \!\!  \textbf{end}  \\

\hlinewd{1.4pt}
\end{tabular}

}

\end{table}

In order to validate the multipath characterization, a scenario with $L = 3$ waves is simulated. The first, second and third wave parameters are: first ($\phi_1 = 270\degree$, $\theta_1 = 90\degree$, $\tau_1 = 30$~ns), second ($\phi_2 = 225\degree$, $\theta_2 = 60\degree$, $\tau_2 = 20$~ns), and third ($\phi_3 = 315\degree$, $\theta_3 = 120\degree$, $\tau_3 = 40$~ns), respectively. For a frequency range $f \nolinebreak \in [58,62]$~GHz with $K = 200$ frequency samples, a toric array with $R = 0.25$~m, $\rho = 0.125$~m, $P = 720$, $M = 300$ is chosen. Fig. \ref{estimacion_multipath} shows the whole process for the multipath estimation of the three waves. Each column illustrates the estimation of the $l$-th wave. For the first path (left column), $\widetilde{\mathbf{H}}(\phi, \tau)$ domain is maximum around the pair \{$\phi_1$, $\tau_1$\}. Note that the other two waves are not visible in a dynamic range of 30 dB due to the amplitude difference introduced by the different values of $\theta_l$ (see Fig. \ref{barrido_theta}). According to $\phi_1$ estimation, $\widetilde{\mathbf{H}}(\theta, \tau_{\phi_1})$ is calculated by taking into account the samples distributed through the ring with $\phi = 270\degree$. This domain is maximum for the pair \{$\theta_1$, $\tau_{1,\phi_1}$\}. The appearance of two vertical lines denotes the presence of two additional waves. In this first step, the estimation does not converge correctly for the two additional waves $(l = 2,3)$ because the plane of incidence is not close to the ring at $\phi = 270\degree$.

Given the first path estimation, we can proceed to estimate the second path after subtracting the first path from the signal $H_{p,z}(f)$. After the application of the phase mode transformation, the inverse filter and the 2D-FFT, we arrive at $\widetilde{\mathbf{H}}(\phi, \tau)$. In this case, the two previously hidden waves are now visible (central column). Note that a slight time dispersion, previously stated in Sect.~II.A, is observed since $\theta_l \neq 90\degree$. This dispersion is not particularly high due to the size of the array, where it is satisfied that $\tau_l \gg 2\rho/c$. Generally, the physical size of the array, and therefore $\rho$, decreases for higher frequencies, which implies a lower dispersion. Nonetheless, the estimation of $\tau_2$ is performed from $\tau_{2,\phi_2}$, thus avoiding the uncertainty of the dispersion. Given $\phi_2 = 225\degree$, $\widetilde{\mathbf{H}}(\theta, \tau_{\phi_2})$ points out the values of $\theta_2$ and $\tau_{2,\phi_2}$. No spectral amplitude is found around $\tau = 30$~ns since the first wave ($l = 1$) has been previously subtracted, while the third wave ($l = 3$) appears around $\tau = 40$ ns, although with high artifact amplitude in the elevation domain. This is because the azimuth estimation plane in the central panel ($\phi_2 = 225\degree$) is far from the plane where the third wave is located ($\phi_3 = 315\degree$). Finally, the third path (right column) is estimated in the same way as in the two previous cases, thus obtaining the trio of values formed by ($\tau_3, \phi_3, \theta_3$).


\section{\label{sec:Conclusions} Conclusions}

This work proposes a method for the joint 3D DoA (azimuth $\phi$, elevation $\theta$) and  time of arrival ($\tau)$ characterization of communication channels. The method is based on the use of frequency-invariant (wideband) beamformer, originally developed for circular arrays and now extended to toric arrays. Through a phase-mode expansion of the channel frequency response in a toric array, this response can be approximated by a function that only depends on the ToA and the DoA coincident with the plane where the circle is located. By taking advantage of the torus definition,  a closed surface formed by the Cartesian product of two circles, several arrays can be defined in multiple planes. Therefore, under an optimal distribution of spatial sampling in the torus, it is possible to use the same sampling to define circular arrays in several planes. Thus, an optimal choice of the samples generates multiple beamformers which are able to estimate the triplet \{$\phi, \theta, \tau$\}.

A parametric analysis of the variables involved has been performed, showing the optimal working range of these variables. Multiple simulations have been carried out for frequencies in the mm-wave range ($30$ GHz and $60$ GHz) providing good results regarding the 3D propagation channel characterization in both, single-path and multipath scenarios. 

Given the trend in the industry to move into the mm-wave frequency range and the growing number of propagation scenarios for cellular communications, the present method arises as a powerful tool to fully characterize these 3D environments. In addition, future research includes the selection of samples from the toric array based on oblique planes passing through the center of the torus. These planes generate circles, known as Villarceau circles, which define new circular arrays in relation to those already shown in this work, and which could improve the response of the frequency-invariant beamformers.

\ifCLASSOPTIONcaptionsoff
  \newpage
\fi

\end{document}